\documentclass[sn-mathphys-num]{sn-jnl}

\usepackage{graphicx}%
\usepackage{subcaption}%
\usepackage{multirow}%
\usepackage{amsmath,amssymb,amsfonts}%
\usepackage{amsthm}%
\usepackage{mathrsfs}%
\usepackage[title]{appendix}%
\usepackage{xcolor}%
\usepackage{textcomp}%
\usepackage{manyfoot}%
\usepackage{booktabs}%
\usepackage{algorithm}%
\usepackage{algorithmicx}%
\usepackage{algpseudocode}%
\usepackage{listings}%
\usepackage{longtable}
\usepackage{orcidlink}

\theoremstyle{thmstyleone}%
\theoremstyle{thmstyletwo}%

\theoremstyle{thmstylethree}%

\begin{document}

\title[Surgeons vs. Computer Vision: A comparative analysis on surgical phase recognition capabilities]{Surgeons vs.\ Computer Vision: A comparative analysis on surgical phase recognition capabilities}


\author*[1,2]{\fnm{Marco} \sur{Mezzina}\orcidlink{0009-0005-1245-1727}}\email{marco.mezzina@orsi.be}

\author[1,3]{\fnm{Pieter} \sur{De Backer}\orcidlink{0000-0002-9375-2353}}

\author[4]{\fnm{Tom} \sur{Vercauteren}\orcidlink{0000-0003-1794-0456}}

\author[2]{\fnm{Matthew} \sur{Blaschko}\orcidlink{0000-0002-2640-181X}}

\author[1,3]{\fnm{Alexandre} \sur{Mottrie}\orcidlink{0000-0002-1253-1592}}

\author[2]{\fnm{Tinne} \sur{Tuytelaars}\orcidlink{0000-0003-3307-9723}}

\affil*[1]{\orgname{Orsi Academy}, \orgaddress{\state{Belgium}}}

\affil[2]{\orgdiv{Faculty of Engineering Science}, \orgname{KU Leuven}, \orgaddress{\state{Belgium}}}

\affil[3]{\orgdiv{Department of Urology}, \orgname{OLV Aalst Hospital}, \orgaddress{\state{Belgium}}}

\affil[4]{\orgdiv{School of BMEIS}, \orgname{King's College London}, \orgaddress{\state{UK}}}


\abstract{\textbf{Purpose}: Automated Surgical Phase Recognition (SPR) uses Artificial Intelligence (AI) to segment the surgical workflow into its key events, functioning as a building block for efficient video review, surgical education as well as skill assessment. Previous research has focused on short and linear surgical procedures and has not explored if temporal context influences experts' ability to better classify surgical phases. This research addresses these gaps, focusing on Robot-Assisted Partial Nephrectomy (RAPN) as a highly non-linear procedure. \textbf{Methods}: Urologists of varying expertise were grouped and tasked to indicate the surgical phase for RAPN on both single frames and video snippets using a custom-made web platform. Participants reported their confidence levels and the visual landmarks used in their decision-making. AI architectures without and with temporal context as trained and benchmarked on the Cholec80 dataset were subsequently trained on this RAPN dataset. \textbf{Results}: Video snippets and presence of specific visual landmarks improved phase classification accuracy across all groups. Surgeons displayed high confidence in their classifications and outperformed novices, who struggled discriminating phases. The performance of the AI models is comparable to the surgeons in the survey, with improvements when temporal context was incorporated in both cases. \textbf{Conclusion}: SPR is an inherently complex task for expert surgeons and computer vision, where both perform equally well when given the same context. Performance increases when temporal information is provided. Surgical tools and organs form the key landmarks for human interpretation and are expected to shape the future of automated SPR.}


\maketitle

\section{Introduction}

Over the past few years the relevance of surgical video recordings has grown, becoming an essential asset across several surgical application domains \cite{cheikh2023evolution}. Recordings are a valuable resource for education and training, enabling observation, retrospective analysis, and remote instruction \cite{mota2018video,schlick2020video}, while also providing objective documentation for investigating surgical techniques and skills that impact patient outcomes \cite{balvardi2022association}. To fully harness the potential of surgical videos, the definition and assessment of surgical phases are crucial, as they enable standardized indexing of procedures, thereby streamlining video review and analysis. With the integration of Artificial Intelligence (AI) in surgery, the computer-assisted classification of surgical events has become more practicable \cite{garrow2021machine}. Surgical Phase Recognition (SPR) automates the assessment of surgical phases by using AI to segment the workflow of a surgery into its key phases.

Deep Learning methods have been previously applied \cite{twinanda2016endonet,jin2017sv,czempiel2021opera} with significant improvements in the performance of SPR classification systems \cite{rivoir2024pitfalls}. However, these methods have only been evaluated for a limited variety of linear and short surgical procedures such as laparoscopic cholecystectomy and cataract surgeries \cite{twinanda2016endonet,zisimopoulos2018deepphase}. Each surgery is unique, with distinct phases, and as the surgical complexity increases, SPR becomes more difficult. Analysing intricate procedures that introduce new challenges—such as longer durations, non-linear progression, and the need to identify subtle phase transitions—is crucial for developing robust and generalisable SPR models. In addition, there has been little focus on understanding the decision-making process of medical professionals when classifying surgical phases. By integrating the foundational knowledge and the critical insights that shape the experts' judgments, reliable algorithms can be developed to support surgeons in automating SPR.

This study benchmarks automated SPR to human performance in an effort to see how well clinicians recognise different surgical phases on complex non-linear Robot-Assisted Partial Nephrectomy (RAPN) procedures when provided with the same data format as computer vision algorithms. The findings aim to (1) clarify the feasibility of SPR for intricate procedures using established AI pipelines, (2) identify the performance gap between clinicians and automated systems, and (3) gain insights into the decision-making process of humans while classifying surgical phases.

\section{Materials and Methods}

\subsection{Datasets}
\subsubsection{Cholec80}

The Cholec80 dataset \cite{twinanda2016endonet} consists of 80 surgical videos of laparoscopic cholecystectomy procedures recorded at 25 Frames Per Second (FPS) with a resolution of either 854x480 or 1920x1080 pixels. Videos last 38.6$\pm$17.1 minutes and include contributions from 13 different surgeons. Annotations are provided as phase timestamps (25 FPS); binary labels for tool presence (1 FPS) are also included but not used in this study. The surgical workflow of a cholecystectomy procedure is categorised into 7 distinct surgical phases, defined by a senior surgeon. Every phase and its duration is listed in Table \ref{table:combined_phases_duration}.

\subsubsection{Robot-Assisted Partial Nephrectomy}
The RAPN dataset used in the survey and for AI development is an extension of previous research of De Backer et al.\ \cite{de2023surgical} and consists of 143 full length surgical recordings of transperitoneal RAPN procedures lasting 122.3$\pm$45.6 minutes, collected under institutional review board approval (OLV Hospital protocol code B12201941209, Ghent University Hospital protocol code B6702020000442). All videos are recorded at 30 FPS, with a resolution of 1280x720 or 1920x1080 pixels. 15 distinct phases are annotated, identical to the methodology of \cite{de2023surgical}, but with removal of the last phase 'Operation Finished' due to irrelevancy in this study. Table \ref{table:combined_phases_duration} reports phase durations.

\begin{table}[t]
\caption{Cholec80 and RAPN phases average durations, which are proposed as mean $\pm$ standard deviation (in minutes). Both dataset exhibit a class imbalance trend.}
\centering
\begin{tabular*}{\textwidth}{@{\extracolsep{\fill}}ll ll}
\toprule
\multicolumn{2}{c}{Cholec80} & \multicolumn{2}{c}{RAPN} \\
\cmidrule(lr){1-2} \cmidrule(lr){3-4}
Phase & Duration (m) & Phase & Duration (m) \\
\midrule
Preparation                    & 2.1 $\pm$ 1.6   & Port Insertion and Surgical Access  & 9.2 $\pm$ 9.7  \\
Calot triangle dissection       & 15.9 $\pm$ 9  & Colon (and Spleen) Mobilization     & 14.2 $\pm$ 10.3  \\
Clipping and cutting            & 2.8 $\pm$ 2.5  & General Hilar Control               & 13.7 $\pm$ 11.7  \\
Gallbladder dissection          & 14.3 $\pm$ 9.2  & Selective Hilar Control             & 12.3 $\pm$ 11.5  \\
Gallbladder packaging           & 1.6  $\pm$ 0.9   & Kidney Mobilization                 & 23.8 $\pm$ 17.8 \\
Cleaning and coagulation        & 3.0 $\pm$ 2.8  & Tumor Identification                & 16.9 $\pm$ 15.7  \\
Gallbladder retraction          & 1.4  $\pm$ 0.9   & Hilar Clamping                      & 2.0  $\pm$ 4.7   \\
---                             & ---            & Tumor Excision                      & 12.1 $\pm$ 16.9  \\
---                             & ---            & Inner Renorrhaphy                   & 9.6 $\pm$ 7.5  \\
---                             & ---            & Hilar Unclamping                    & 4.0 $\pm$ 8.8  \\
---                             & ---            & Outer Renorrhaphy                   & 10.4 $\pm$ 6.6  \\
---                             & ---            & Specimen Retrieval                  & 3.0 $\pm$ 7.7  \\
---                             & ---            & Specimen Removal and Closing        & 5.6 $\pm$ 10.6  \\
---                             & ---            & Retroperitonealization of the Kidney & 6.9 $\pm$ 4.4  \\
---                             & ---            & Instrument Removal                  & 4.2 $\pm$ 3.9 \\
\bottomrule
\end{tabular*}
\label{table:combined_phases_duration}
\end{table}

\subsection{Survey}\label{section:survey_structure}
\subsubsection{Sampling strategy}
Every participant was randomly assigned one RAPN procedure from the dataset when starting the survey and every procedure was assigned only once. The data presented to a participant follows a sampling strategy which aims to select sufficiently diverse and qualitative samples. Additionally, ambiguous sections such as phase transitions and out-of-body sequences were discarded.

Sampling is done temporally by selecting timestamps. The probability of selecting a timestamp within phase $p$ follows a Normal distribution $\mathcal{N}(\mu_p,\sigma_p)$, where $\mu_p$ represents the midpoint of $p$, and $\sigma_p$ is 0.1 times the duration of $p$. Timestamps within 30 seconds of each other were discarded to avoid redundancy. In addition, we did not fetch frames close to phase transitions, excluding the timestamps from the 5\textit{th} and 95\textit{th} percentiles of each phase. For every sampled timestamp, we extract both the single frame at that stamp and a video snippet of the 10 seconds preceding the single frame. For every procedure, 17 timestamps are sampled thus resulting in 17 single images and 17 video snippets. We ensured that at least one timestamp was sampled for each phase. However, as there are 17 samples and only 15 phases, we oversampled the longer phases (exceeding 10 minutes). This choice was made to account for potential variability, considered that longer phases may contain multiple distinct and representative segments.

Despite the structured approach, ensuring high-quality visual content for sampled timestamps remains challenging due to the inherent randomness of the sampling process, which may result in noisy frames and snippets, affected by blurring from rapid camera movements, excessive bleeding and fogging during suturing actions.

\subsubsection{Survey structure}
The survey was deployed as a custom-made website, freely accessible to medical students, residents, fellows, and consultants in Urology. Volunteers were recruited in person and through posts on social media platforms. The survey is as a series of consecutive dynamic web pages structured as follows:
\begin{enumerate}
    \item \textbf{Introduction}: The homepage welcomes participants, explaining the study's overall goal, its estimated completion time (30 minutes) and how the collected non-personal data, obtained with Informed Consent, are treated. The study starts with users sharing information about their profession (Medical Student, Resident, Fellow or Consultant), their expertise level (Viewers, Bedside Assistant, Supervised Console Surgeon or Independent Console Surgeon) and the number of assisted RAPN cases ($<$50, 50-100, 100-200, 200-500, $>$500). Upon continuing, the web platform assigns a RAPN procedure to the participant and fetches the related single frames and video snippets. 
    \item \textbf{Phase Description}: Respecting good practices to avoid cognitive biases, as explained in \cite{ellis2018so}, a validated textual description of the 15 RAPN phases, developed by surgeons for surgeons, is presented \cite{farinha2023international}. Participants are instructed to carefully review it before proceeding.
    \item \textbf{Questions - Single Frames}: The 17 sampled single frames are shuffled and presented one-by-one to the participant \cite{ellis2018so}. For each image, the user is asked to classify the estimated surgical phase, by selecting it from the full list of phases, and rate its confidence level on a discrete scale from 1 (not confident - random choice) to 5 (very confident). The response time per question is also tracked.
    \item \textbf{Feedback - Single Frames}: The participant is asked to rate the difficulty of the task, indicating a numerical score on a discrete scale from 1 (difficult survey) to 5 (easy survey). Additionally, they are asked to type in a text box the visual landmarks used in their decision-making process.
    \item \textbf{Questions - Video Snippets}: Now the 17 video snippets of the same procedure are shuffled, independently from the frames in the previous set of questions to avoid the users relying on previous answers, and presented again one-by-one to the user. The confidence rating is asked and the response time is tracked.
    \item \textbf{Feedback - Video Snippets}: Once again, we inquire for user feedback, likewise to single frames. Then, a mandatory yes/no question asks if the video snippet length was sufficient to complete the task, followed by an optional question asking a feedback about the overall survey experience.
    \item \textbf{Final Section}: The survey ends with a thank you note summarising the performance of the participant per section. A web link is provided for those who wish to stay informed about the study results guaranteeing transparency and anonymity.
\end{enumerate}

We opted to not collect personal data to promote broader participation, as these were unnecessary for the assessment of users' performance. 

The decision of using 17 questions per section was a trade-off between getting sufficient data for the study while minimising completion time and ensuring a high level of concentration throughout the survey. For the same reasons, we decided to not include a time-consuming preparatory tutorial showing sample frames and video snippets to human raters prior the classification task. The tutorial, even if it might be considered best practice to ensure a fairer comparison with AI algorithms, was deemed unnecessary since the survey targeted individuals familiar with RAPN procedures—those who had observed, assisted, or performed such complex surgeries.

\subsection{AI Pipeline}\label{section:ai_pipeline}
Parallel to the survey, we investigated how established SPR algorithms compare to the performance of clinicians for the task of phase recognition in RAPN mirroring the study design. We trained a ResNet50 \cite{he2016deep} and a ResNet50-Long Short-Term Memory (LSTM) for single frames and video snippets End-To-End classification to automate SPR, as performed by Jin et al. \cite{jin2017sv}. Moreover, we trained TeCNO \cite{czempiel2020tecno}, with a dilation factor of 1, to extend the comparison between well-known SPR AI pipelines. No spatial labels, such as tool and organ segmentation masks, were used as additional input for the tested frameworks due to their unavailability in the current dataset version.

We benchmarked the AI models by first replicating results on the publicly available Cholec80 dataset and then applying the same experimental setup to the novel RAPN dataset. In both cases, frames were sampled at 1 FPS and resized to 224x224. Black borders in some RAPN recordings, caused by the resolution of the surgical device, were removed. Further preprocessing was equal for both Cholec80 and the RAPN dataset. To enhance model robustness and generalisation, data augmentation was applied with a probability 50\%, including horizontal flipping, rotation (up to 15 degrees), blurring, brightness adjustment, and random erasing. Normalization was performed using ImageNet mean and standard deviation for Cholec80, while custom values were computed for RAPN. The Cholec80 dataset was partitioned according to the established splits by Jin et al. \cite{jin2017sv} and Czempiel et al. \cite{czempiel2020tecno}. For the RAPN dataset, we simulated varying levels of medical expertise by training models on subsets containing 25\%, 50\%, 75\%, and 100\% of the data. These experiments are submitted to \textit{5}-fold cross-validation, with full-length videos assigned exclusively to either the training or validation set.

The input of the LSTM module in both experiments consists of the ResNet50 feature embeddings computed on a buffer of $N$ consecutive frames, which corresponds to the last $N$ seconds of the video due to 1 FPS sampling. The LSTM state is not propagated from one $N$-seconds buffer to the next, but it is propagated only within the buffer. We evaluated two buffer durations, testing $N$ with values of 10 and 60 seconds to capture varying temporal contexts. The 10-seconds buffer ensures comparability with the participants' clips, while the 60-seconds buffer extends the temporal context for AI models, even though it is not directly comparable to humans. During training, there was no overlap between the buffers so every frame was part of only one buffer. The ResNet50 backbone was initialised with ImageNet weights, while the LSTM and TeCNO were initialised with random weights. All models were trained for 30 epochs.

\section{Results}

\subsection{Study Participants}

\begin{figure}[b]\centering
\subfloat[Profession]{\label{fig:professions_histogram}\includegraphics[width=.33\linewidth]{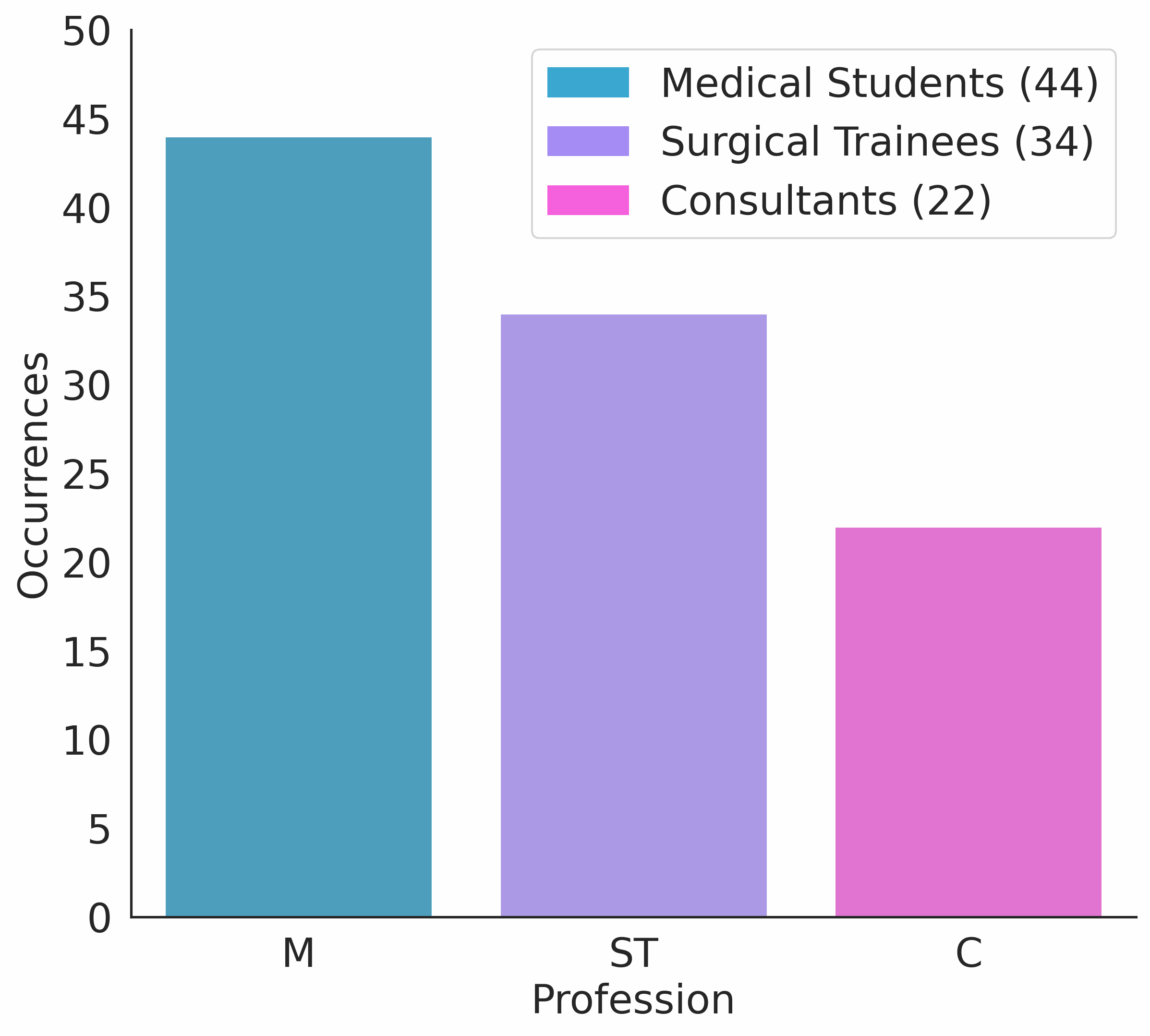}}
\subfloat[Expertise level]{\label{fig:experience_histogram}\includegraphics[width=.33\linewidth]{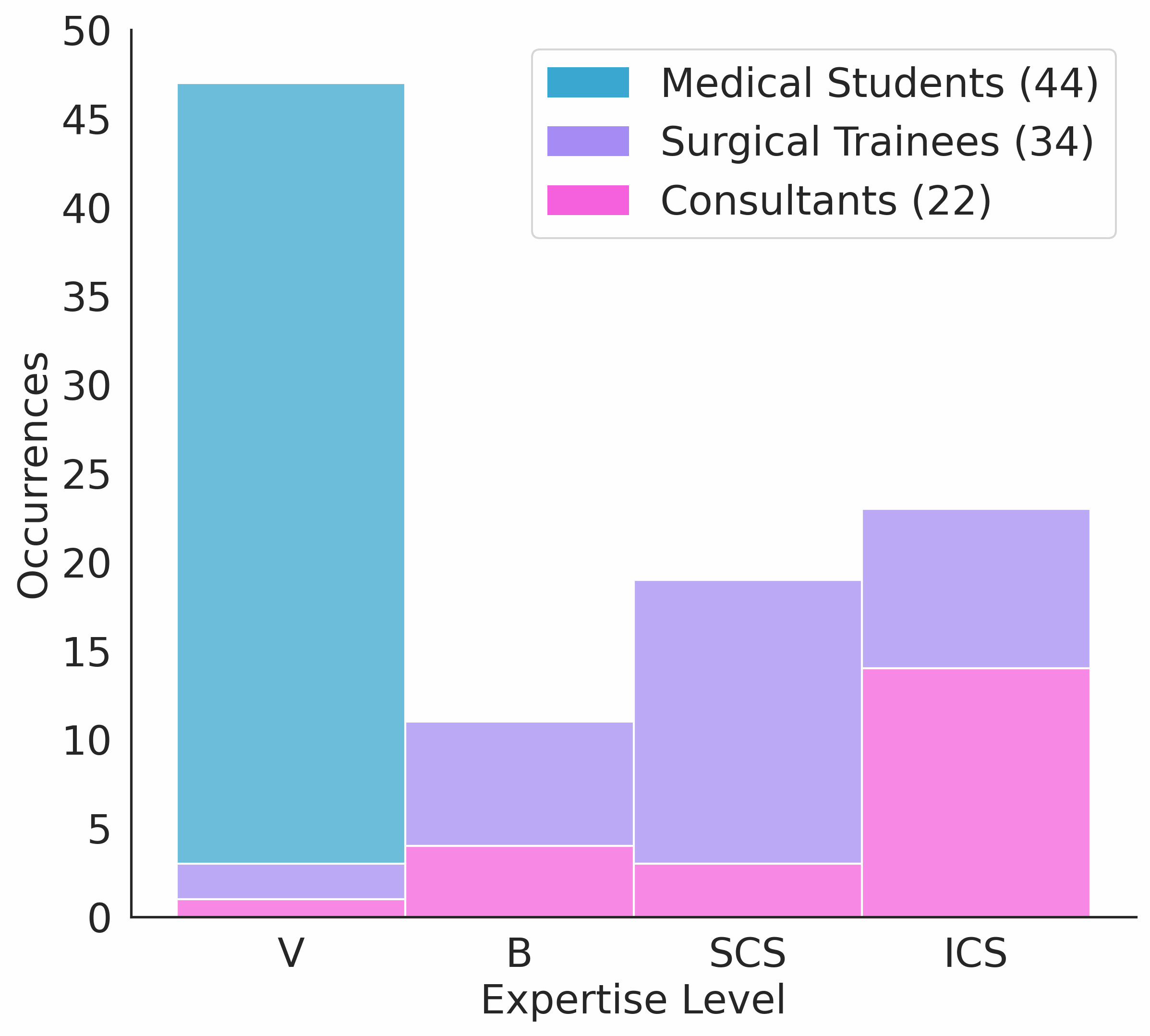}}
\subfloat[Assisted cases]{\label{fig:assisted_cases}\includegraphics[width=.33\linewidth]{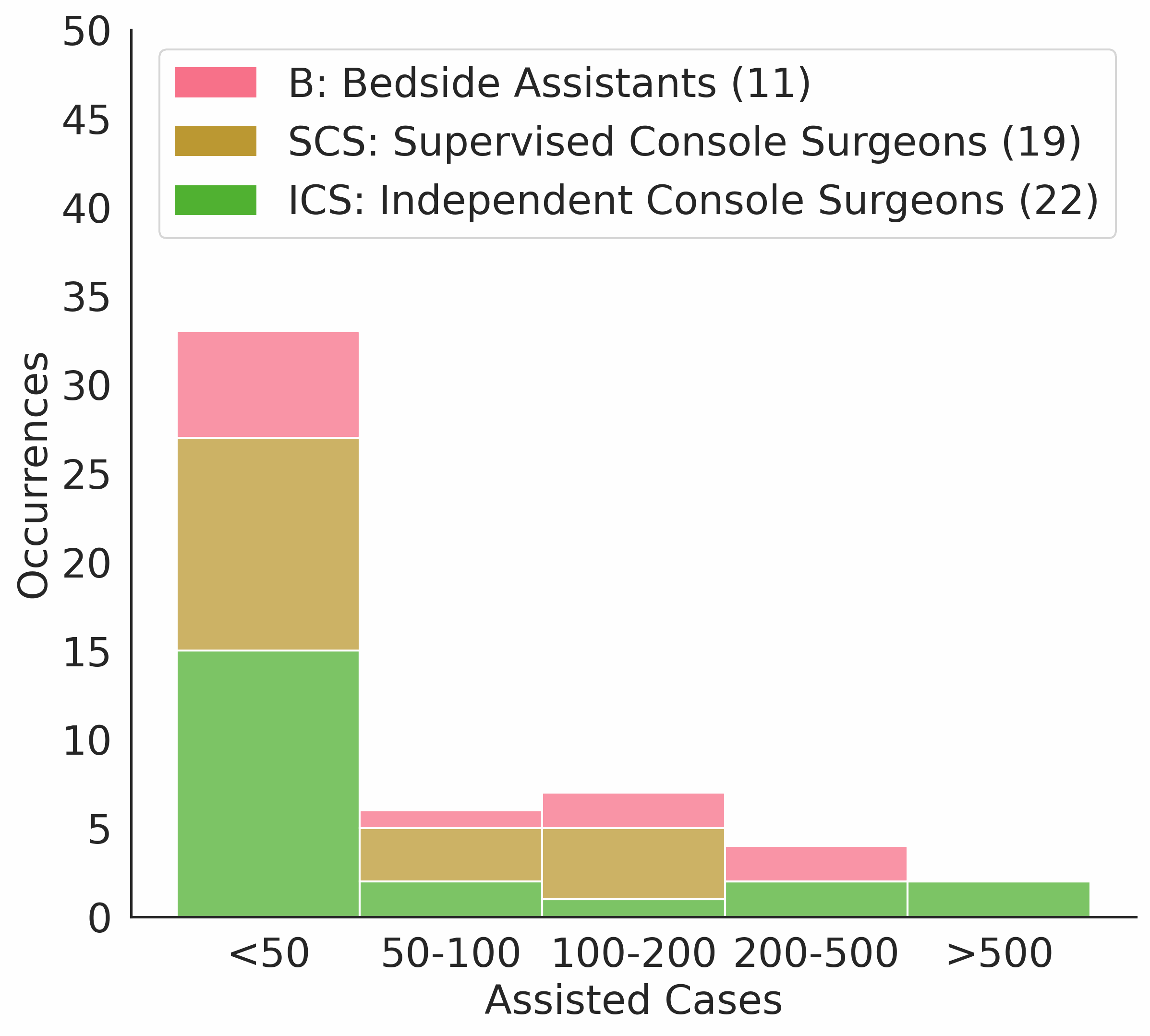}}
\caption{Survey Participants. Figure \ref{fig:professions_histogram} - M: Medical Students, ST: Surgical Trainees, C: Consultants. Figure \ref{fig:experience_histogram} - V: Viewers, B: Bedside Assistants, SCS: Supervised Console Surgeons, ICS: Independent Console Surgeons. Figure \ref{fig:assisted_cases} - Number of urology procedures in which B, SCS and ICS have been involved during their careers.}
\label{fig:survey_participants}
\end{figure}

100 participants, including medical students, residents, fellows and consultants filled in the survey over a 6 month period.
We merged 24 residents and 10 fellows into one single group labelled \textit{Surgical Trainees}, given their similar levels of expertise. Moreover, their performance did not differ significantly from one another.
We linked the profession of the participants (Figure \ref{fig:professions_histogram}) with their level of expertise (Figure \ref{fig:experience_histogram}). Finally, we related the expertise level to the number of urological cases actively assisted inside the operating room (Figure \ref{fig:assisted_cases}). 

\subsection{Single Frames Versus Video Snippets}\label{section:single_frames_vs_video_snippets}

\begin{figure}[b]
     \centering
     \begin{subfigure}[b]{0.6\textwidth}
         \centering
         \includegraphics[scale=0.054]{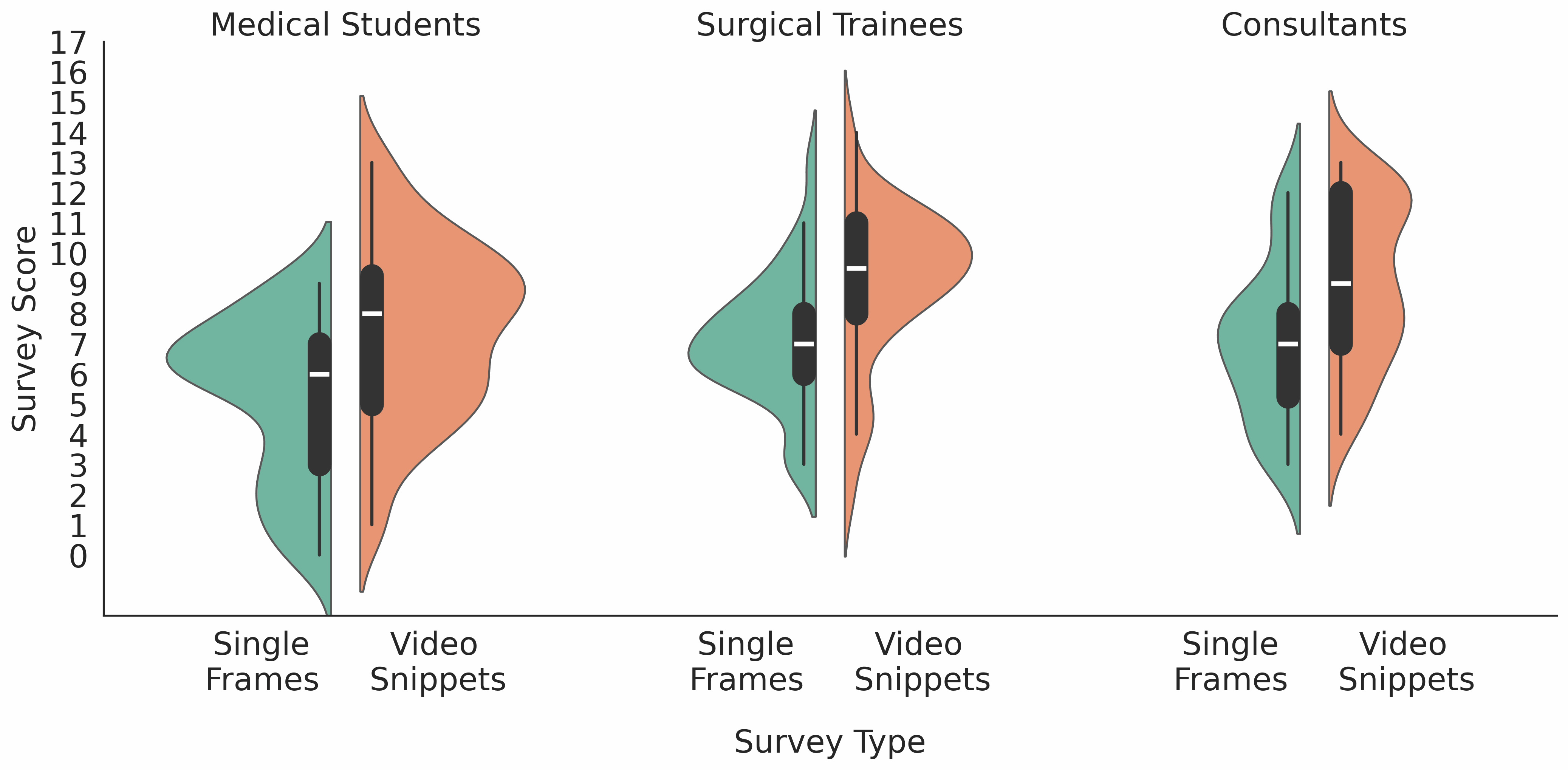}
         \caption{Scores across professions. Densities are proportional to the number of observations.}
         \label{fig:per_group_performance}
     \end{subfigure}
     \hfill
     \begin{subfigure}[b]{0.3\textwidth}
         \centering
         \includegraphics[scale=0.1]{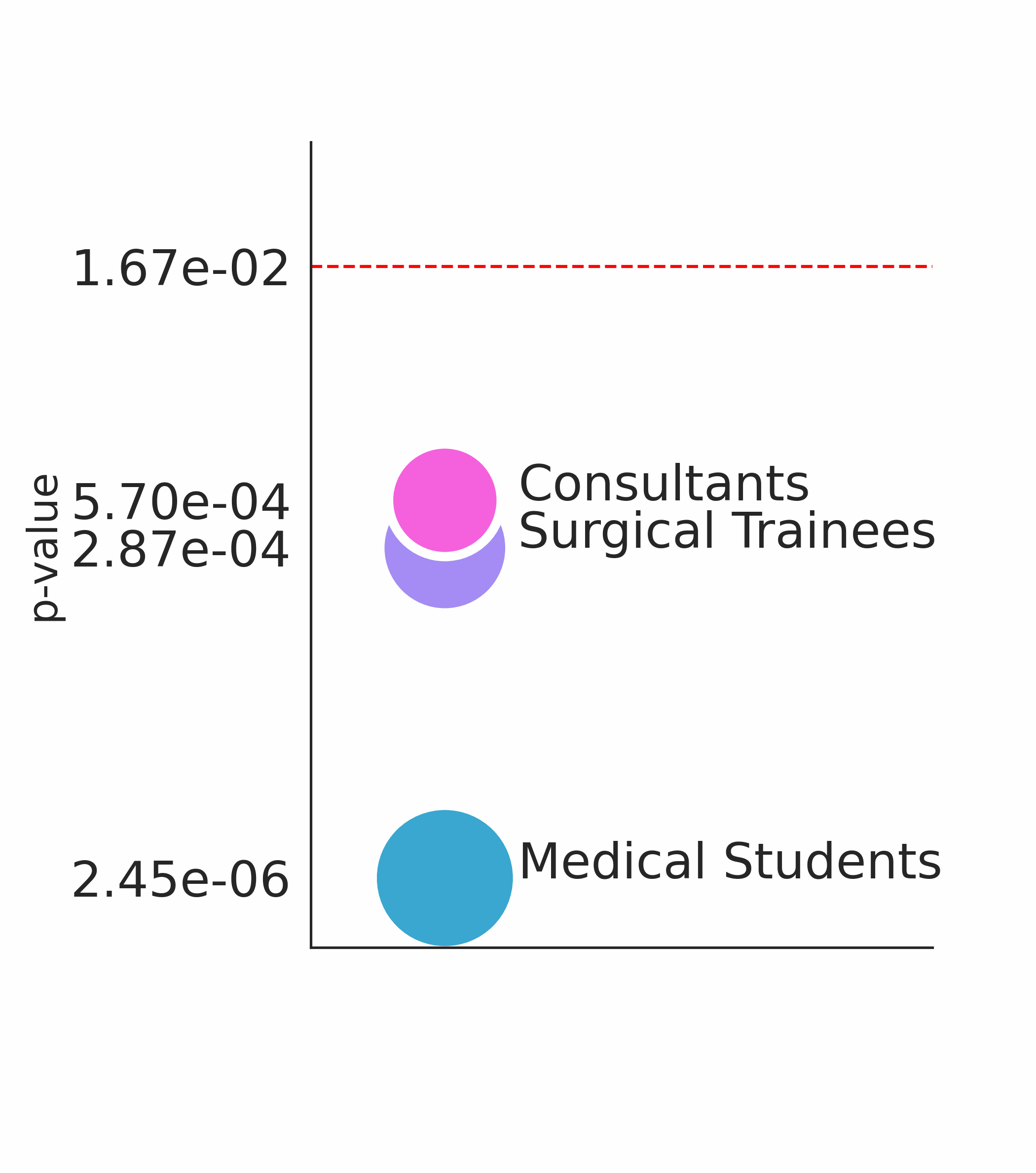}
        \caption{Wilcoxon Signed-Rank test results.}
        \label{fig:wilcoxon}
     \end{subfigure}
    \caption{Null Hypothesis $H_0$ rejected for all groups professions.}
    \label{fig:global_performance}
\end{figure}

To test the hypotheses that video snippets have an impact on SPR accuracy compared to single frames in RAPN procedures, we selected the Wilcoxon Signed-Rank test \cite{conover1999practical, 2020SciPy-NMeth} as a paired difference test between single frames and video snippets scores. The following hypothesis were formulated:
\begin{itemize}
    \item Null Hypothesis $H_0$: There is no significant difference in SPR accuracy between single frames and video snippets. The median difference in accuracy between the two conditions is zero.
    \item Alternative Hypothesis $H_1$: There is a significant difference in SPR accuracy between single frames and video snippets.
\end{itemize}
We applied the Bonferroni correction to counteract the multiple comparisons problem \cite{bonferroni1936teoria}. Thus, we divided the $\alpha=0.05$ threshold per the number of comparisons. Figure \ref{fig:global_performance} summarises users' scores (Figure \ref{fig:per_group_performance}) and the p-values resulting from the Wilcoxon Signed-Rank test (Figure \ref{fig:wilcoxon}) per profession. The Null Hypothesis was rejected for all groups and the profession significantly influenced the p-values. No significant relationship was observed between the time that participants spent answering the questions and their performance.

Table \ref{table:ranked_landmarks_split} lists the occurrences of clustered visual landmarks denoted by participants with free text to classify surgical phases in RAPN procedures, ranked in descending order. The high amount of \textit{Movements \& Gestures} attributions highlights the value of having temporal context.

\begin{table}[t]
\caption{Occurrences of landmarks denoted by participants to classify surgical phases in RAPN procedures, ranked in descending order. $^1$ Set composed by text words: \textit{movement of tools}, \textit{moving tools}, \textit{gestures} and \textit{actions done by instruments}. $^2$ Set composed by text words: \textit{organs/tissues state/deformations} and \textit{damage to tissues}.}
\centering
\begin{tabular*}{\textwidth}{@{\extracolsep{\fill}}l c c l c c}
\toprule%
\multicolumn{3}{c}{\textbf{Surgical Tools}} & \multicolumn{3}{c}{\textbf{Organs}} \\
\cmidrule(lr){1-3} \cmidrule(lr){4-6}
\textbf{Landmark} & \shortstack{\textbf{Single} \\ \textbf{Frames}} & \shortstack{\textbf{Video}\\\textbf{Snippets}} & \textbf{Landmark} & \shortstack{\textbf{Single} \\ \textbf{Frames}} & \shortstack{\textbf{Video}\\\textbf{Snippets}} \\
\midrule
\textit{Surgical Tools} & 71 & 92 & \textit{Anatomic Structures} & 63 & 81 \\
\textit{Movements \& Gestures}$^1$ & - & 48 & \textit{Tissues Status}$^2$ & 13 & 9 \\
\textit{Sutures \& Sutures Wires} & 16 & 8 & \textit{Tumor} & 5 & 7 \\
\textit{Vessel Loops} & 9 & 3 & \textit{Blood \& Bleeding} & 8 & 3 \\
\textit{Clips} & 5 & 4 & \textit{Kidney} & 5 & 6 \\
\textit{Needle} & 5 & 3 & \textit{Colon} & 2 & 2 \\
\textit{Clamps} & 3 & 2 & \textit{Fat} & 2 & 2 \\
\textit{Bulldog Clamps} & 2 & 2 & \textit{Artery \& Veins} & 3 & 2 \\

\textit{Trocars} & - & 1 & \textit{Spleen} & 1 & 1 \\
\textit{Hemostatic Agent} & - & 1 & \textit{Endobag} & 1 & 1 \\
\botrule
\end{tabular*}
\label{table:ranked_landmarks_split}
\end{table}

To assess the appropriateness of video snippets length, participants were asked whether 10 seconds were sufficient to classify phases or not. While 79 preferred longer snippets and 21 found 10 seconds adequate, performance did not differ significantly between the groups, though consultants who favoured shorter snippets exhibited higher confidence. However, feedback from participants who preferred longer snippets highlighted two key points. First, one participant noted that "\textit{some videos don't represent certain phases and they should last several seconds}". While the ideal snippet length was not specified, incorporating long-term temporal context might help resolve ambiguities in the complex surgical workflow of RAPN surgeries, which result from varying visual patterns in patient anatomy, diverse surgical techniques, and phase sequences broken into smaller actions. Second, another participant commented that "\textit{videos should not be longer but should capture more relevant moments of the phases}". Although no specific criteria to select informative frames was suggested, the feedback indicates that identifying key moments that summarise phases may be a valuable research direction to improve classification while minimising video snippets length.

\subsection{Novices Versus Experts}

To measure the spread of participants' answers, we computed the $L_{\frac{1}{2}}$ norm for each row of the confusion matrix, which yields values in the range $[1,k]$ (with \textit{k}=15 as the total number of RAPN phases), where 1 indicates a one-hot distribution, whereas 15 denotes a uniform distribution, implying maximal spread across all phases. Next, we averaged the rows outcomes and normalised the final score to lie within $[0, 1]$, with the goal of providing a quantitative measure that summarises the uncertainty in decision-making during phase classification.

Table \ref{tab:survey_performance} summarises the average Accuracy, weighted F1-Score and $L_{\frac{1}{2}}$ norm for all categories. Medical Students were most confused, as compared to Surgical Trainees and Consultants. The phases '\textit{Hilar Control}', '\textit{Tumor Excision}', and '\textit{Hilar Clamping}' benefited most from video snippets, as we believe the temporal context clarified the sequence of actions. Notably, all groups generally struggled with the classification of short duration phases.

\begin{figure}[t]
    \centering
    \includegraphics[scale=0.065]{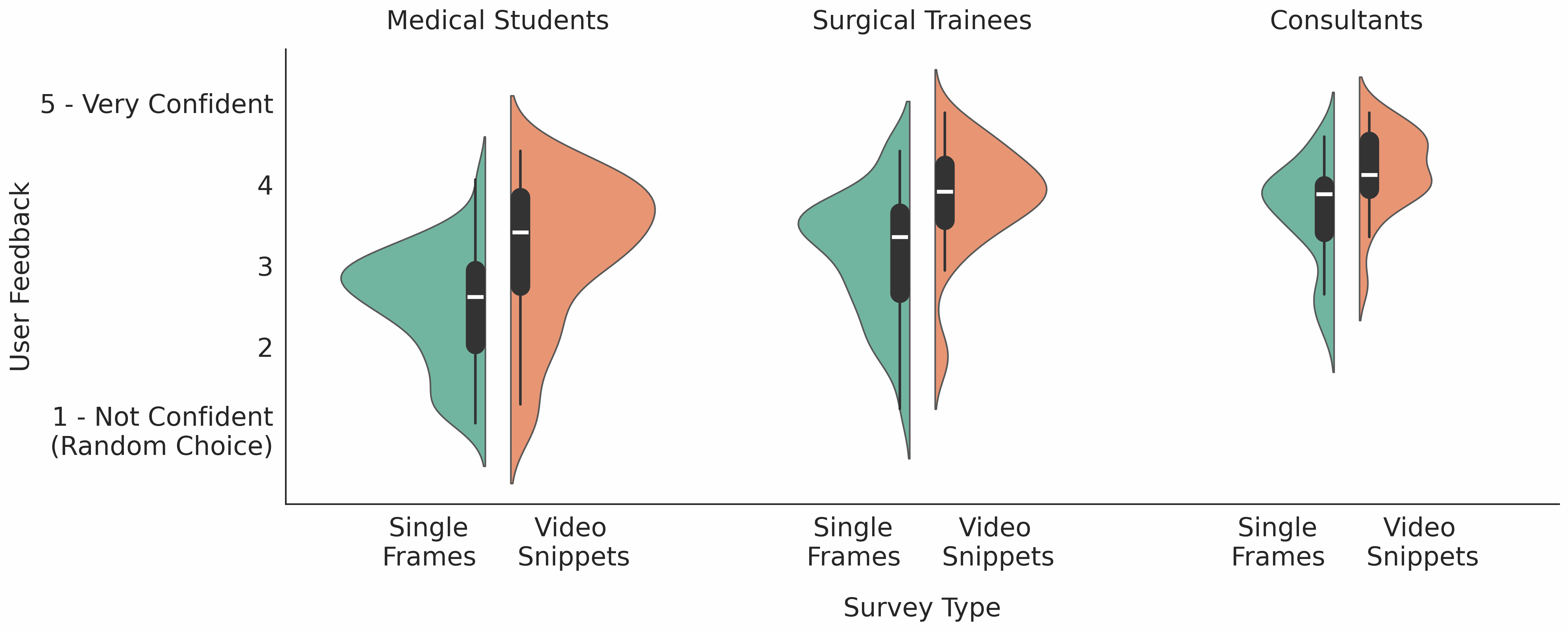}
    \caption{Confidence level during the survey across professions. Densities are proportional to the number of samples.}
    \label{fig:per_group_confidence}
\end{figure}

Figure \ref{fig:per_group_confidence} provides additional insights by illustrating the average confidence level in answers per profession. Consultants exhibited the highest confidence. Confidence is a subjective metric, and we computed the Spearman's rank correlation coefficient \cite{spearman1961proof, 2020SciPy-NMeth} between the accuracy and the confidence for each group. The results showed no significant correlation, indicating that subjective confidence has minimal influence on objective classification task performance.

\begin{figure}[t]
    \centering
    \includegraphics[scale=0.065]{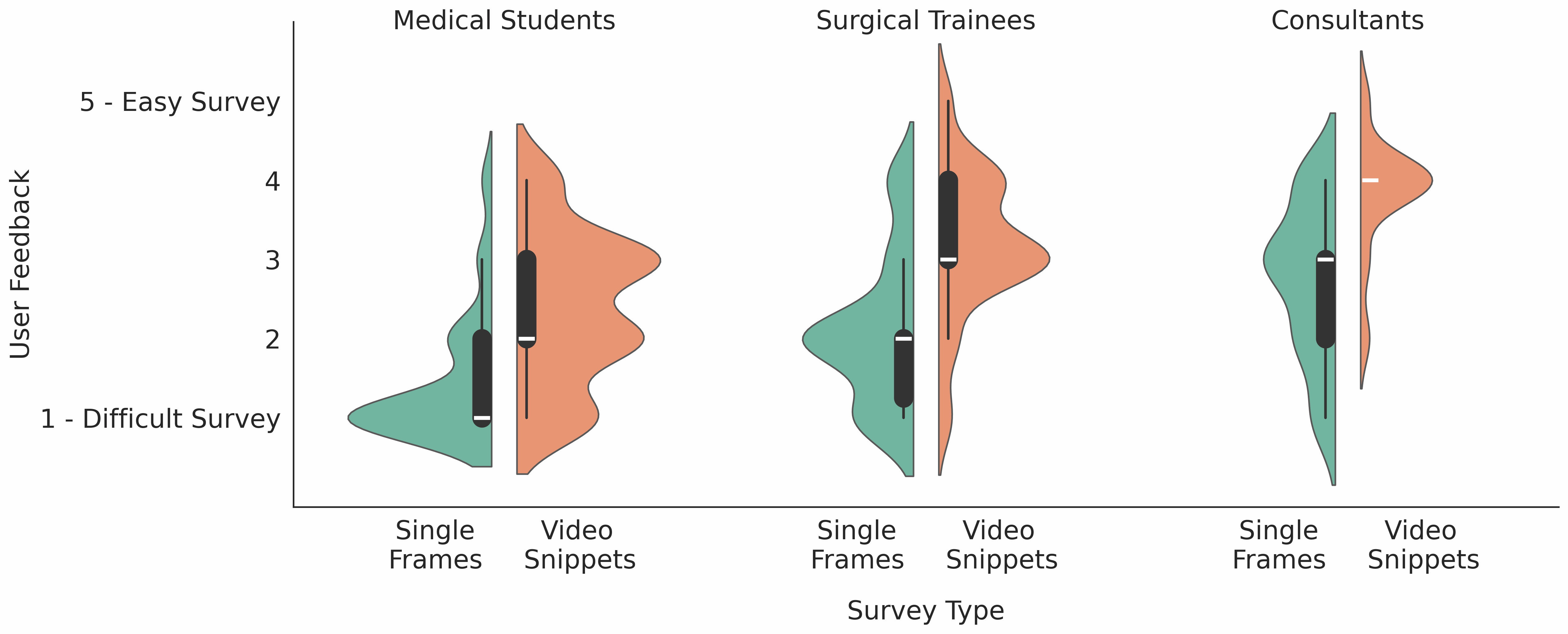}
    \caption{Perception of survey sections difficulty across professions. Densities are proportional to the number of samples.}
    \label{fig:per_group_survey_difficulty_level}
\end{figure}

Figure \ref{fig:per_group_survey_difficulty_level} shows the average perceived task complexity across professions per section. Similar to the confidence findings, Consultants generally found the task to be simpler when working with video snippets. Participants had mixed opinions regarding the difficulty of the task with single frames, especially when compared to the average confidence reported for each individual question. We also calculated Spearman's rank correlation coefficient between the accuracy and the perceived task complexity per group, finding no significant correlation.

\begin{table}[b]
\caption{Participants performance}
\centering
\begin{tabular*}{\textwidth}{@{\extracolsep{\fill}}c cc cc cc}
    \toprule
    \multicolumn{7}{c}{\textbf{Participants Performance}} \\
    \midrule
    \multicolumn{1}{c}{} & \multicolumn{2}{c}{\textbf{\shortstack{Average\\Accuracy (\%)}}} & \multicolumn{2}{c}{\textbf{\shortstack{Average Weighted\\F1-Score (\%)}}} & \multicolumn{2}{c}{\textbf{\shortstack{Average\\L$\frac{1}{2}$ Norm ([0,1])}}} \\
    \cmidrule(lr){2-3} \cmidrule(lr){4-5} \cmidrule(lr){6-7}
    \textbf{Group} & \textbf{\shortstack{Single\\Frames}} & \textbf{\shortstack{Video\\Snippets}} & \textbf{\shortstack{Single\\Frames}} & \textbf{\shortstack{Video\\Snippets}} & \textbf{\shortstack{Single\\Frames}} & \textbf{\shortstack{Video\\Snippets}} \\
    \midrule
    Medical Students  & 30.3  & 43.6  & 27.9  & 40.9  & 0.54 & 0.43 \\
    Surgical Trainees & 41.5  & 52.9  & 36.5  & 47.8  & 0.37 & 0.28 \\
    Consultants       & 41.2  & \textbf{53.2}  & 36.1  & \textbf{48.2}  & 0.33 & \textbf{0.27} \\
    \bottomrule
\end{tabular*}
\label{tab:survey_performance}
\end{table}

\subsection{AI Performance}

We validated ResNet50-LSTM and TeCNO on Cholec80, achieving average accuracies of 82.4\% and 84.9\% respectively, which is comparable to the literature \cite{jin2017sv,czempiel2020tecno}.

Tables \ref{tab:rapn_results} summarise the performance of ResNet50, ResNet50-LSTM and TeCNO for the complex RAPN dataset. The addition of more RAPN labelled data improves classification performance, but the gains decrease as more videos are incorporated. Notably, a longer-temporal context of 60-seconds for ResNet50-LSTM only enhances performance when the full dataset is available, indicating that lengthy buffers require a larger volume of labelled data. Long duration phases benefitted most from 60-seconds buffers, contrarily to short phases who suffered the strategy of training without overlapping clips. Interestingly, the $L\frac{1}{2}$ norm decreased with the extension of the temporal context, suggesting a reduced spread across all phases during classification.

The AI models trained on the complete labelled dataset outperformed consultants on single frames and video snippets classification. However, the generally higher confusion level of AI models may stem from the significantly larger volume of samples they processed compared to participants. In contrast, while consultants experienced a significant 12\% accuracy boost with video snippets, AI models showed a more modest improvement, suggesting the human advantage in precisely localising surgical phases by leveraging prior knowledge. 

In addition, when trained with 25\% of the data, the AI models surpassed medical students in classifying frames and videos, demonstrating comparable confusion levels with 10-second clips. Their accuracy also aligned closely with that of Surgical Trainees, though the models exhibited a higher confusion level for single frames.

TeCNO outperformed the ResNet50-LSTM model by extending its field of view to analyse the entire surgical workflow. To maintain a fair comparison with human raters, we deliberately avoided propagating the LSTM state across consecutive \textit{N}-second buffers. Instead, state propagation was restricted to within individual buffers.

Both human raters and AI models struggled with '\textit{Hilar Control},' '\textit{Specimen Retrieval},' '\textit{Hilar Unclamping},' '\textit{Specimen Removal},' and '\textit{Instrument Removal}' phases.

\begin{table}[b]
\caption{RAPN \textit{5}-fold cross validation under different data scarcity scenarios. Results are reported as mean $\pm$ standard error of the 5-folds.}
\centering
\begin{tabular*}{\textwidth}{@{\extracolsep{\fill}}l c c c c c c}
    \toprule
        \multicolumn{6}{c}{\textbf{RAPN - ResNet50}} \\
    \midrule
    \multicolumn{1}{c}{\textbf{Metric}} & \multicolumn{1}{c}{\textbf{Model}} & \multicolumn{4}{c}{\textbf{Dataset Availability}} \\
    \cmidrule(lr){3-6}
    & & 25\% & 50\% & 75\% & 100\% \\
    \midrule
    \multirow{4}{*}{Accuracy (\%)} 
    & No LSTM & 45.9$\pm$0.02 & 50.7$\pm$0.01 & 54.8$\pm$0.01 & 54.9$\pm$0.01 \\
    & LSTM (10s) & 50.0$\pm$0.03 & 56.1$\pm$0.01 & 58.7$\pm$0.01 & 58.8$\pm$0.01 \\
    & LSTM (60s) & 48.1$\pm$0.04 & 52.6$\pm$0.02 & 58.7$\pm$0.01 & 60.6$\pm$0.01 \\
    & TeCNO & 53.1$\pm$0.02 & 60.0$\pm$0.02 & \textbf{61.6$\pm$0.01} & 61.4$\pm$0.02 \\
    \midrule
    \multirow{4}{*}{\shortstack{Weighted \\ F1-Score (\%)}} 
    & No LSTM & 45.8$\pm$0.02 & 51.2$\pm$0.01 & 55.5$\pm$0.01 & 55.6$\pm$0.01 \\
    & LSTM (10s) & 47.6$\pm$0.04 & 55.6$\pm$0.02 & 58.6$\pm$0.01 & 58.8$\pm$0.01 \\
    & LSTM (60s) & 42.2$\pm$0.04 & 46.3$\pm$0.03 & 52.9$\pm$0.01 & 55.5$\pm$0.02 \\
    & TeCNO & 49.3$\pm$0.02 & 57.7$\pm$0.01 & \textbf{60.5$\pm$0.01} &60.1$\pm$0.01 \\
    \midrule
    \multirow{4}{*}{L$\frac{1}{2}$ Norm ([0,1])} 
    & No LSTM & 0.53 & 0.51 & 0.50 & 0.49 \\
    & LSTM (10s) & 0.41 & 0.42 & 0.41 & 0.43 \\
    & LSTM (60s) & 0.27 & 0.24 & 0.25 & \textbf{0.23} \\
    & TeCNO & 0.3 & 0.34 & 0.34 & 0.38 \\
    \bottomrule
\end{tabular*}
\label{tab:rapn_results}
\end{table}

\section{Conclusions}

In this study, we benchmark automated SPR against the performance of urologists with varying levels of expertise to evaluate the ability of clinicians to classify different surgical phases using the same data format provided to computer vision algorithms. For this purpose, we deployed a custom web-based survey.

We demonstrate that short video snippets improved classification accuracy across all professional groups and influenced the selection of visual landmarks. Consultants achieved the highest scores, but exhibited an overconfidence trend. AI models outperformed consultants in accuracy; however, they also demonstrated higher levels of confusion, likely due to the complexity of the RAPN surgical workflow. The results suggest promising potential for AI, though it still falls short of inflated expectations of classifying with a minimal error rate, especially in contexts where experts themselves face challenges. 

Our study reflects the inherent common challenges of automated SPR, but it has limitations. Some sampled single frames and video snippets contained visual noise and redundant gestures that could distract from key actions, while variations in patient anatomy might cause similar phases to appear differently. We believe that incorporating longer-term temporal context, exceeding 60-seconds, might help mitigating such ambiguities by capturing a broader view of the surgical workflow. In addition, we tested only one video snippet length to balance participant engagement with survey feasibility.

In the future, once labels become available, we plan to investigate the impact of multi-task strategies that incorporate tool and organ segmentation masks on SPR during the classification stage. Additionally, we aim to explore the link between urology expertise and label quality. While consultants performed best, their high confidence levels raise questions about efficient data annotation and mitigating confirmation bias \cite{ellis2018so}. Lastly, we intend to explore the benefits of several minutes temporal context for RAPN surgeries.

\backmatter

\bmhead{Supplementary information}
The article has accompanying supplementary file.

\bmhead{Acknowledgements}
The authors thank all participants and Junior Orsi for their contribution and invaluable availability. We also thank Giovanni Cacciamani, Riccardo Campi, Karel Decaestecker, Charles Van Praet and Joris Vangeneugden for their contribution with videos donation and efforts spreading the survey. Finally, the authors thank Jente Simoens, Jasper Hofman and Wouter Bogaert for their support.

\subsection*{Declarations}

\bmhead{Funding}
This research was funded by the following grants: Baekeland grant of Flanders Innovation \& Entrepreneurship (VLAIO), grant number HBC.2020.2252 (Pieter De Backer) and HBC.2023.0156 (Marco Mezzina).

\bmhead{Ethics approval}
The study was approved by KU Leuven’s Privacy and Ethics platform (PRET) and was reviewed by the Social and Societal Ethics Committee (SMEC) of KU Leuven with approval number G-2024-7885-R2(MIN). The procedures used in this study adhere to the tenets of the Declaration of Helsinki.

\bmhead{Consent to Participate and Publish}
Informed consent, including consent for publication, was digitally obtained from all participants.

\bmhead{Data Availability} 
The RAPN dataset is available upon request from the corresponding author for non-commercial use. The data are not publicly available, as the open sourcing of all videos for general usage was not declared in the IRB approval.

\bmhead{Financial \& Non-Financial Conflicts of Interests} 
Tom Vercauteren is co-founder and shareholder of Hypervision Surgical Ltd. The other authors declare no conflicts of interest.


\bibliography{sn-bibliography}

\newpage

\begin{appendices}

\section*{Supplementary Material - Tables} 
\addcontentsline{toc}{section}{Supplementary Material} 
\renewcommand{\thesection}{S\arabic{section}}
\renewcommand{\thetable}{S\arabic{table}}
\renewcommand{\thefigure}{S\arabic{figure}}

\begin{longtable}{|c|p{5.8cm}|}
    \caption{Surgical phases description for RAPN surgery provided to participants at the start of the study. The textual description is the result of validated studies resulted in a protocol developed by surgeons for surgeons.} \\
    \hline
    \textbf{Phase} & \textbf{Description} \\
    \hline
    \endfirsthead

    \hline
    \textbf{Phase} & \textbf{Description} \\
    \hline
    \endhead

    \hline
    \endfoot

    \hline
    \endlastfoot

    \textbf{Port Insertion and Surgical Access} & The endoscopic camera captures the abdominal cavity. Trocars are being placed, instruments are introduced into trocars, and adhesion removal is performed. \\
    \hline
    \textbf{Colon Mobilization} & Identification of the mesentric line and incision of line of Toldt. \\
    \hline
    \textbf{General Hilar Control} & Identification, dissection and isolation of the main renal artery. A vessel loop is placed around the main renal artery. \\
    \hline
    \textbf{Selective Hilar Control} & Identification, dissection and isolation of secondary renal arteries. Vessel loops are placed around the secondary renal arteries. \\
    \hline
    \textbf{Kidney Mobilization} & Incisions of Gerota's fascia to access the kidney. The kidney surface is freed from perinephric fat. \\
    \hline
    \textbf{Tumor Identification} & Localisation and delineation of the tumor. The ultrasound probe can be employed for localisation, while coagulation can be used for tumor marking. \\
    \hline
    \textbf{Hilar Clamping} & Bulldog clamp clamps the main renal artery and/or secondary arteries. \\
    \hline
    \textbf{Tumor Excision} & Resection of the tumor by separation from the renal parenchyma. Not to be confused with Tumor Identification. \\
    \hline
    \textbf{Specimen Retrieval and Deposition} & The tumor mass is placed to a convenient location for later retrieval. \\
    \hline
    \textbf{Inner Renorrhaphy} & Suturing of the inner layers of the kidney including running medullary sutures and repair of the urinary collecting system. \\
    \hline
    \textbf{Hilar Unclamping} & The bulldog clamp is removed from the main renal artery and/or secondary arteries. \\
    \hline
    \textbf{Outer Renorrhaphy} & Suturing/approximation of the kidney cortex. Non-absorbable clips are repeatedly attached to the ends of the suture wire to perform the sliding Hem-o-lok technique. \\
    \hline
    \textbf{Specimen Removal} & The tumor is placed inside an endobag for removal from the patient's body. \\
    \hline
    \textbf{Retroperitonealization of The Kidney} & The peritoneum is closed by performing parietal peritoneum suturing. \\
    \hline
    \textbf{Instrument Removal} & Vessel loops, bulldog clamps, robotic instruments, suture wires are removed from the patient's abdomen. \\
    \hline
\end{longtable}

\begin{longtable}{|p{0.95\linewidth}|}
\caption{Summary of user feedback per user group on the overall survey experience. The table includes only responses from participants who provided it.} \\
\hline
\multicolumn{1}{|c|}{\textbf{Medical Students}} \\ 
\hline
\begin{itemize}
    \item Very straightforward survey, well done and good luck!
    \item Too many unclear clips and images due to gauze/hemostatics, making phase recognition difficult.
    \item Impossible to answer questions with frames; some videos were not representative of phases.
    \item Some videos only showed needle movement, making identification hard.
    \item Suggested adding surgical anatomical videos for recognizing structures in robotic surgery.
    \item Difficult survey. Some videos don't represent certain phases and they should last several seconds. However, it's an interesting survey and it's not possible define how many seconds the videos should last.
    \item Pretty hard for medical students.
    \item Nice survey, but videos had desaturated colors.
    \item Videos should be longer, but defining the right length is tricky.
    \item Single image of clamp didn’t clarify clamping/unclamping.
    \item Difficult frames and short videos.
    \item Even with short videos, distinguishing phases was possible, but frames alone were insufficient.
    \item Videos made it easier to answer.
    \item Some snippets weren’t representative of the phases.
    \item Some phases, like hilar clamping/unclamping, were difficult to recognize, and no videos represented these phases.
    \item Difficult survey, but video snippets made answering easier.
\end{itemize} \\
\hline
\multicolumn{1}{|c|}{\textbf{Residents}} \\
\hline
\begin{itemize}
    \item Very interesting exercise.
    \item A brilliant way to test and improve knowledge for residents.
    \item Suggested longer videos and avoiding non-steady frames or clips with no specific movement.
    \item Requested feedback on correct answers.
    \item Some questions were problematic: Question 8 had no correct answer (hemostasis phase); Question 12 had no video.
    \item Sometimes pictures alone were misleading and could represent multiple situations.
    \item Very difficult!
\end{itemize} \\
\hline
\multicolumn{1}{|c|}{\textbf{Fellows}} \\
\hline
\begin{itemize}
    \item Difficult to match some steps with videos. For example, removing the vessel loop: Is it unclamping or instrument removal? Another example: showing the suture surface—is it outdoor suturing or unclamping? It seems like checking the suture after unclamping.
\end{itemize} \\
\hline
\multicolumn{1}{|c|}{\textbf{Consultants}} \\
\hline
\begin{itemize}
    \item Suggested adding more detailed descriptions for some steps.
    \item The video should not be longer but should capture more relevant moments of the phases.
    \item Very interesting project.
    \item Not all phases shown were listed in the steps.
    \item Some videos showed “in between phases,” potentially confusing both humans and AI.
    \item Hemostatic agents considered as outer renorraphy? Unclear.
    \item Some phases were “in between,” making answers difficult at times.
    \item Some videos didn’t show the specific part of the case in question.
    \item Would like the study’s purpose better explained.
\end{itemize} \\
\hline
\end{longtable}

\begin{table}[h]
\caption{Cholec80 test set performance reported as mean $\pm$ standard deviation. For the 40/40 split, the standard deviation is computed across all test videos, while for the 40/8/32 split, it is computed across multiple runs.}
\centering
\begin{tabular*}{\textwidth}{@{\extracolsep{\fill}}l c c c}
    \toprule
    \multicolumn{1}{c}{\textbf{Metric}} & \multicolumn{1}{c}{\textbf{Model}} & \multicolumn{2}{c}{\textbf{Dataset Split (train/val/test)}} \\
    \cmidrule(lr){3-4}
    & & 40/40/0 & 40/8/32 \\
    \midrule
    \multirow{4}{*}{\textbf{Accuracy (\%)}} 
    & ResNet50+LSTM\cite{jin2017sv} & 85.3$\pm$7.3 & N/A \\
    & ResNet50+LSTM (Our) & 82.4$\pm$8.1 & N/A \\
    & TeCNO\cite{czempiel2020tecno} & N/A & \textbf{88.6$\pm$0.3} \\
    & TeCNO (Our) & N/A & 84.9$\pm$0.4 \\
    \midrule
    \multirow{4}{*}{\textbf{F1-Score (\%)}} 
    & ResNet50+LSTM\cite{jin2017sv} & 82.1$\pm$7.2 & N/A \\
    & ResNet50+LSTM (Our) & $80.2\pm7.6$ & N/A \\
    & TeCNO\cite{czempiel2020tecno} & N/A & \textbf{83.4$\pm$0.6} \\
    & TeCNO (Our) & N/A & 81.0$\pm$0.7 \\
    \bottomrule
\end{tabular*}
\label{tab:cholec80_results}
\end{table}

\section*{Supplementary Material - Figures}

\begin{figure}[H]
    \centering
    \includegraphics[width=\textwidth]{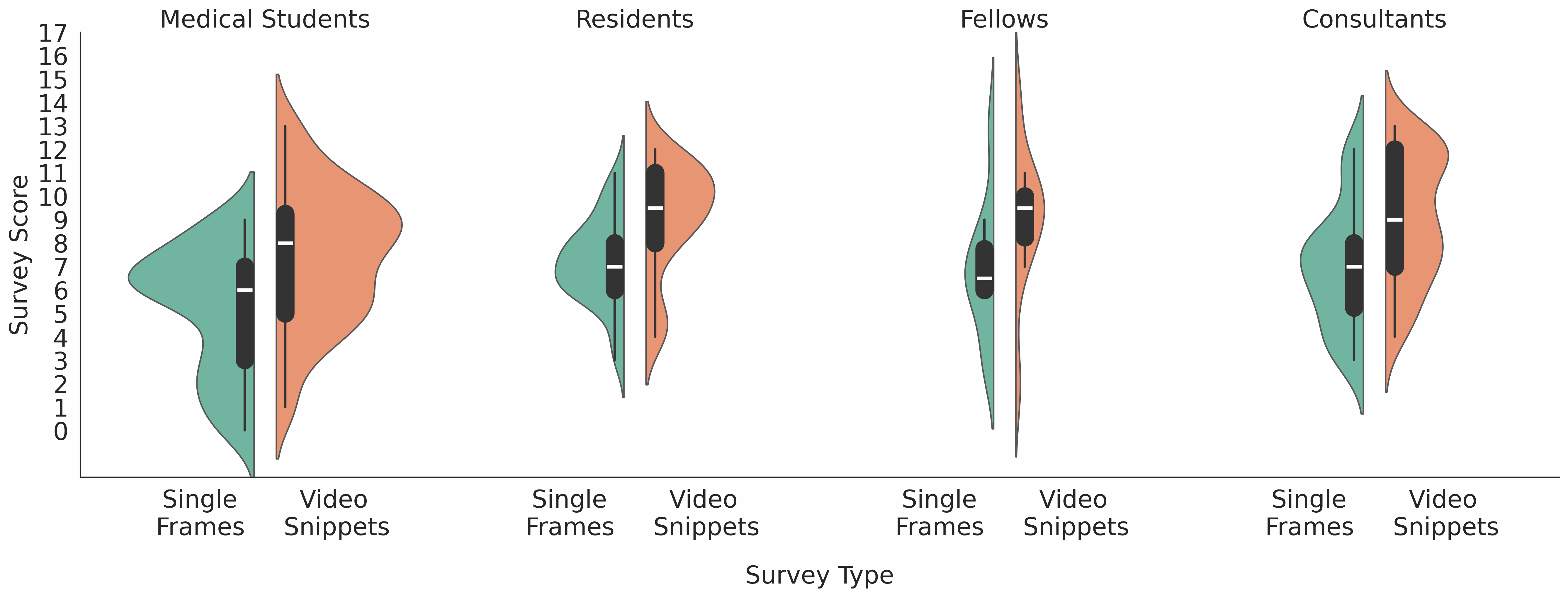}
    \caption{Scores across professions, with Surgical Trainees divided into the original categories of Residents and Fellows. Densities are proportional to the number of observations.}
    \label{fig:original_per_group_performance}
\end{figure}

\begin{figure}[H]
    \centering
    \includegraphics[width=\textwidth]{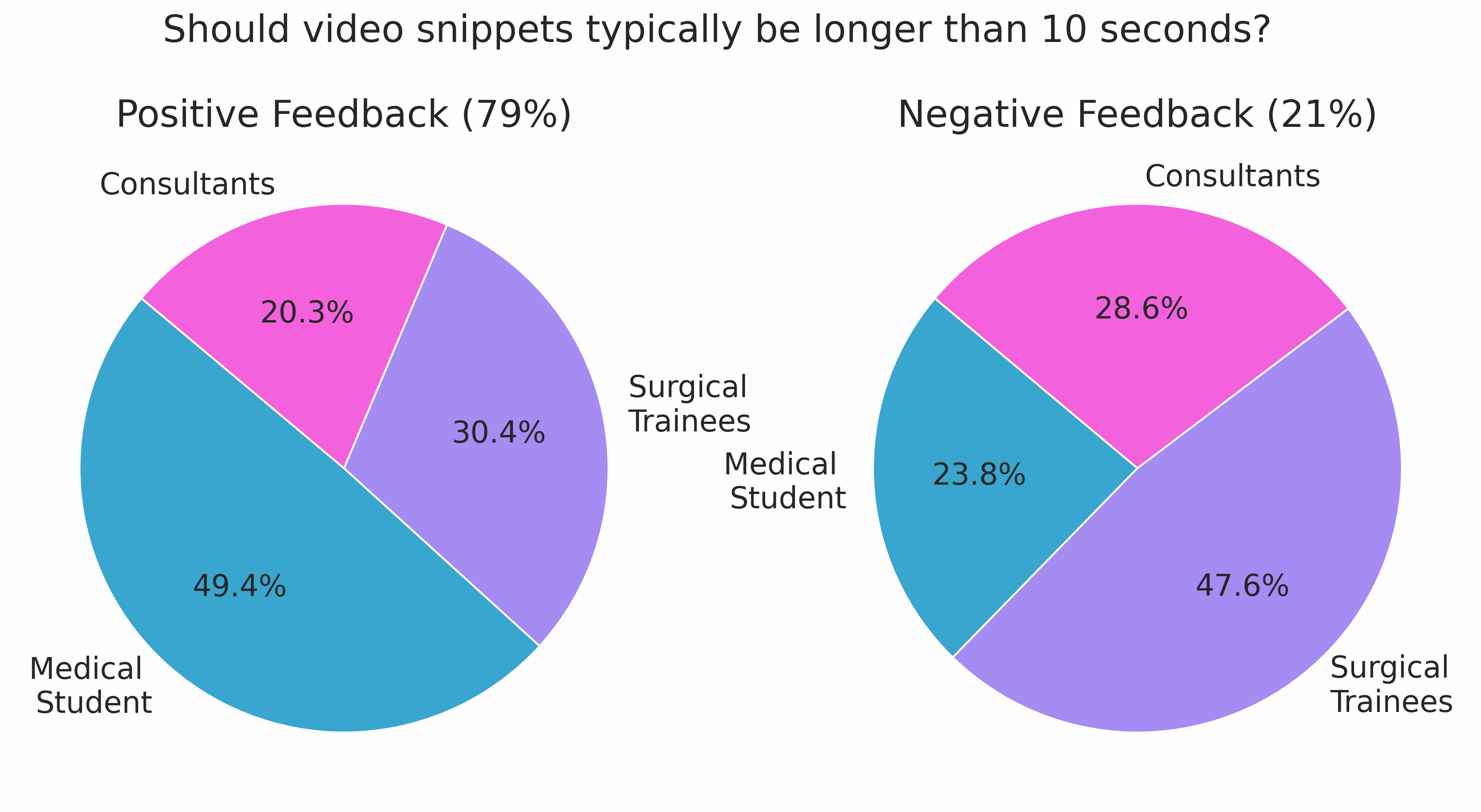}
    \caption{Users feedback on the length of video snippets to solve the classification task.}
    \label{fig:video_lenght_feedback}
\end{figure}

\begin{figure}[H]
     \centering
     \begin{subfigure}[b]{1\textwidth}
        \centering
        \includegraphics[width=1\linewidth]{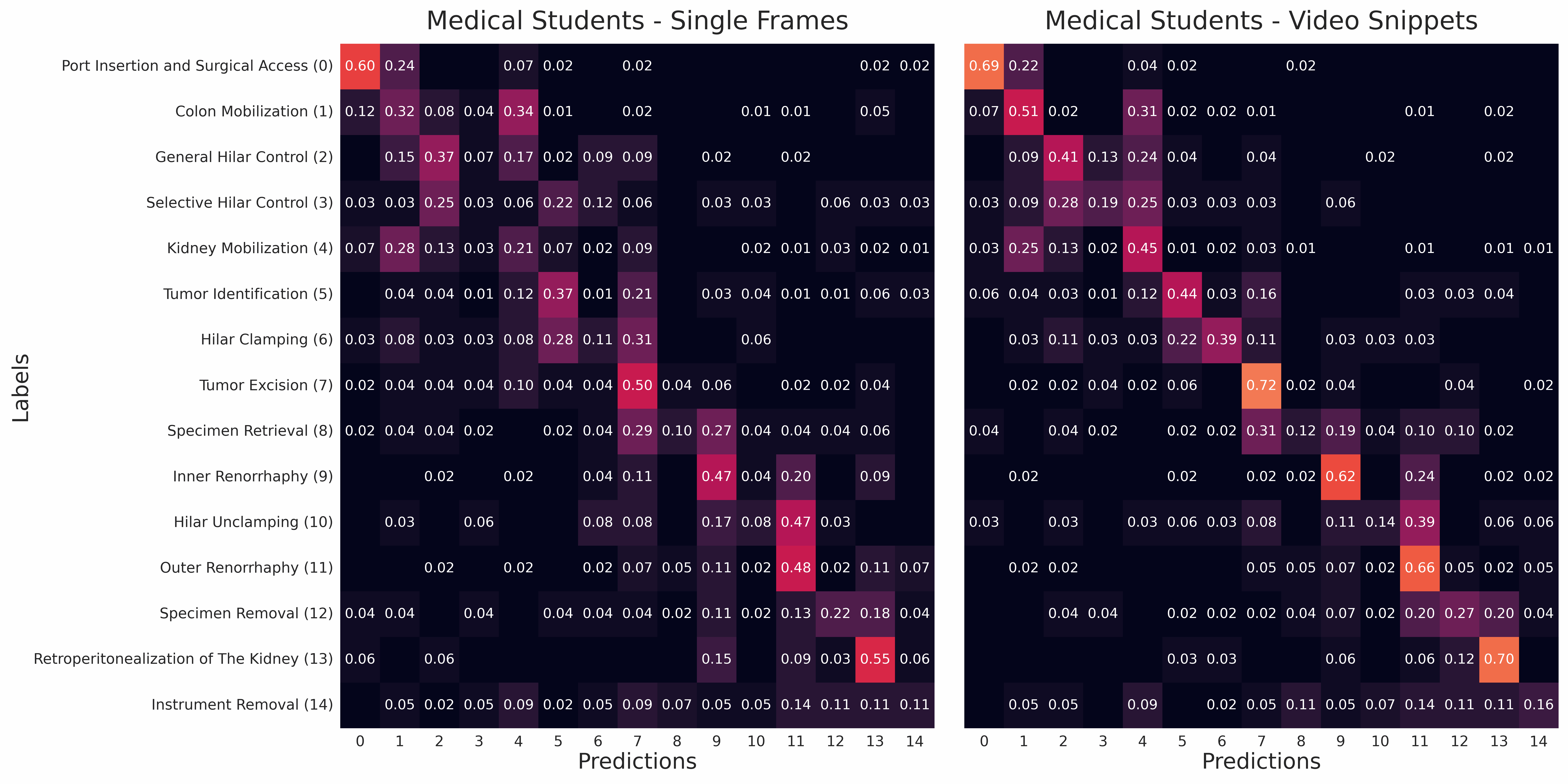}
        \caption{Medical students Performance}
        \label{fig:confusion_matrix_M}
     \end{subfigure}
     \begin{subfigure}[b]{1\textwidth}
        \centering
        \includegraphics[width=1\linewidth]{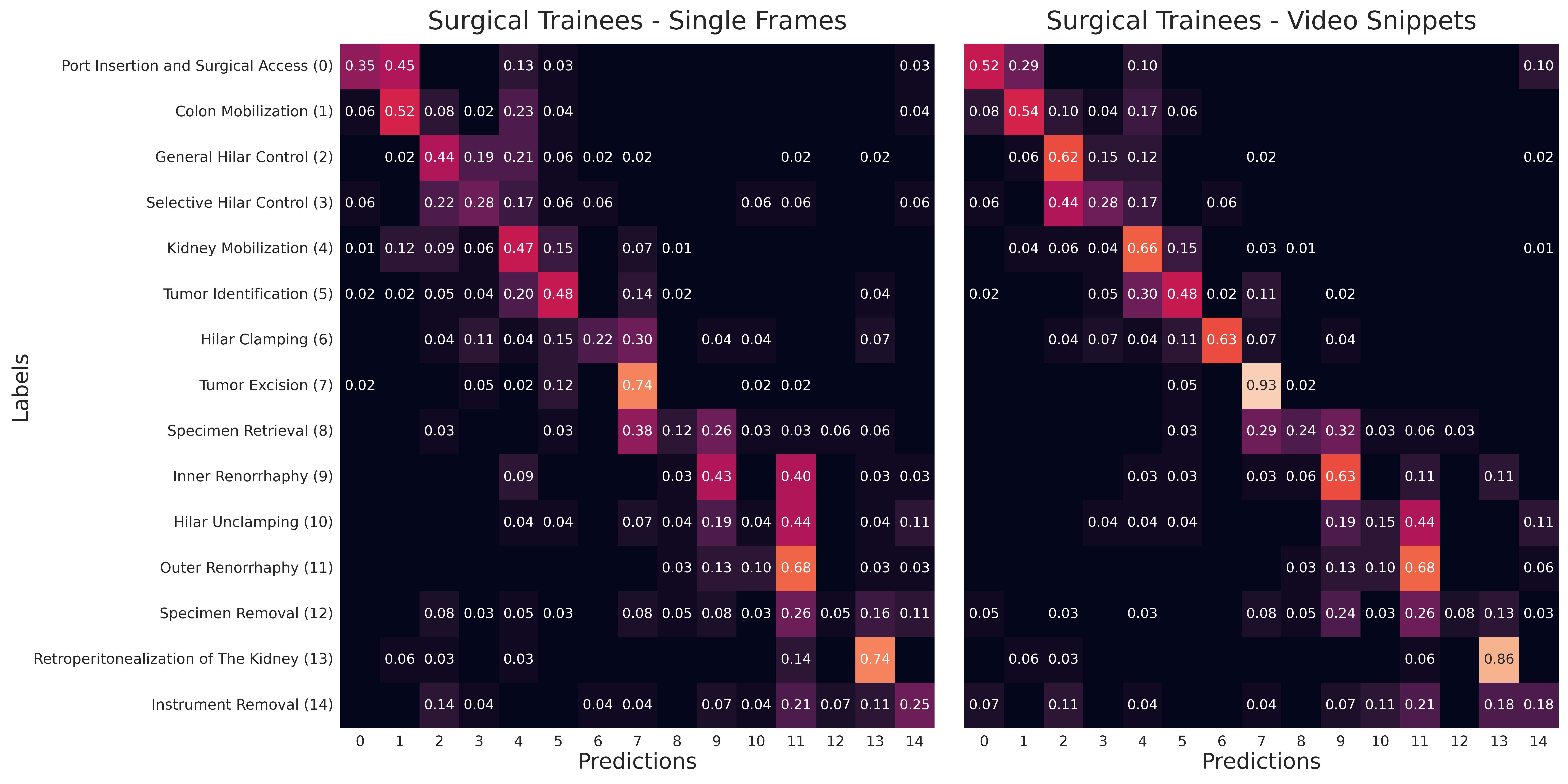}
        \caption{Surgical Trainees Performance}
        \label{fig:confusion_matrix_ST}
     \end{subfigure}
     \begin{subfigure}[b]{1\textwidth}
        \centering
        \includegraphics[width=1\linewidth]{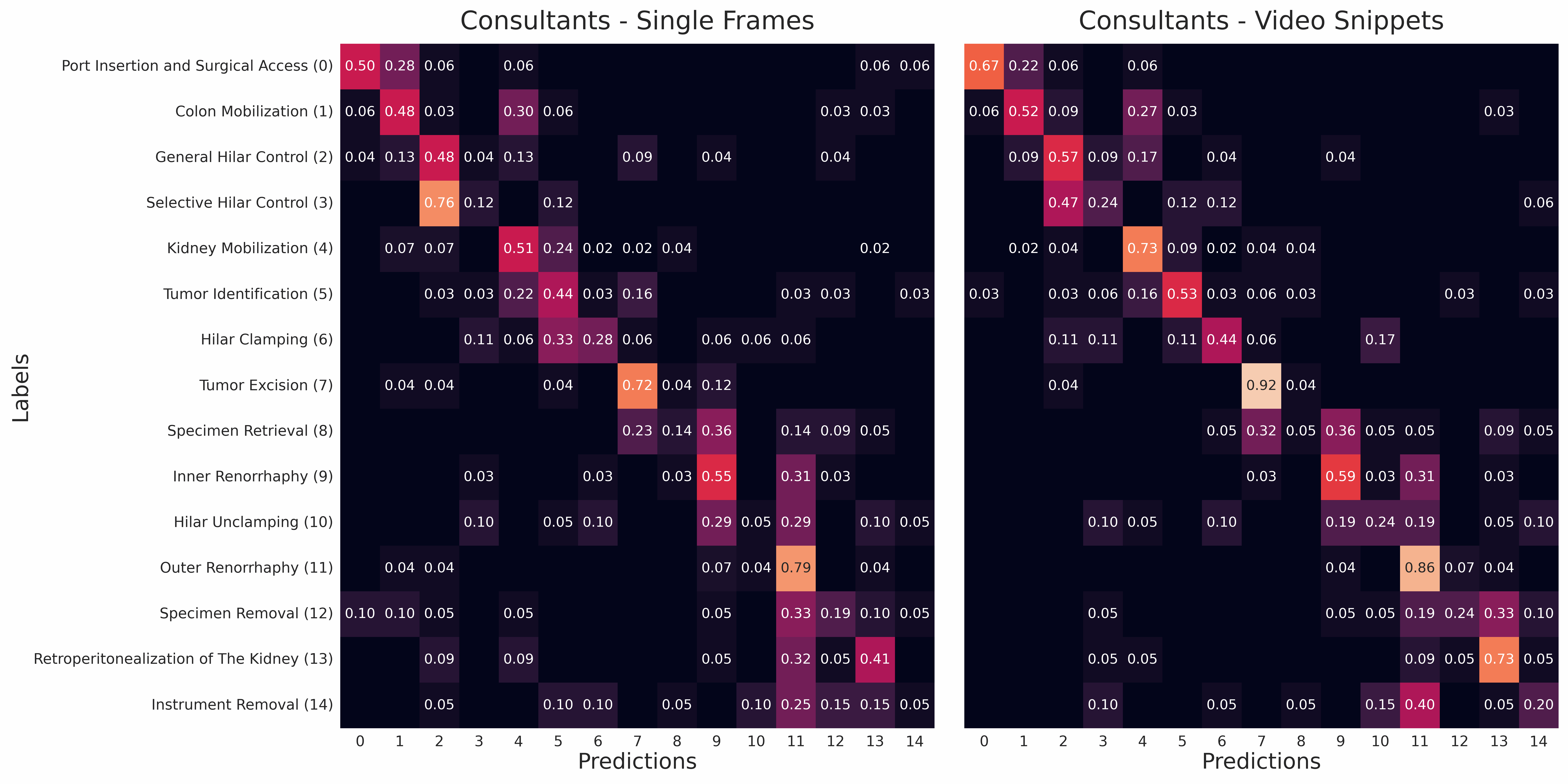}
        \caption{Consultants Performance}
        \label{fig:confusion_matrix_C}
     \end{subfigure}
     \caption{Confusion matrices of survey participants normalised over the predictions}
     \label{fig:confustion_matrices_participants}
\end{figure}

\begin{figure}[H]
     \centering
     \begin{subfigure}[b]{1\textwidth}
        \centering
        \includegraphics[width=1\linewidth]{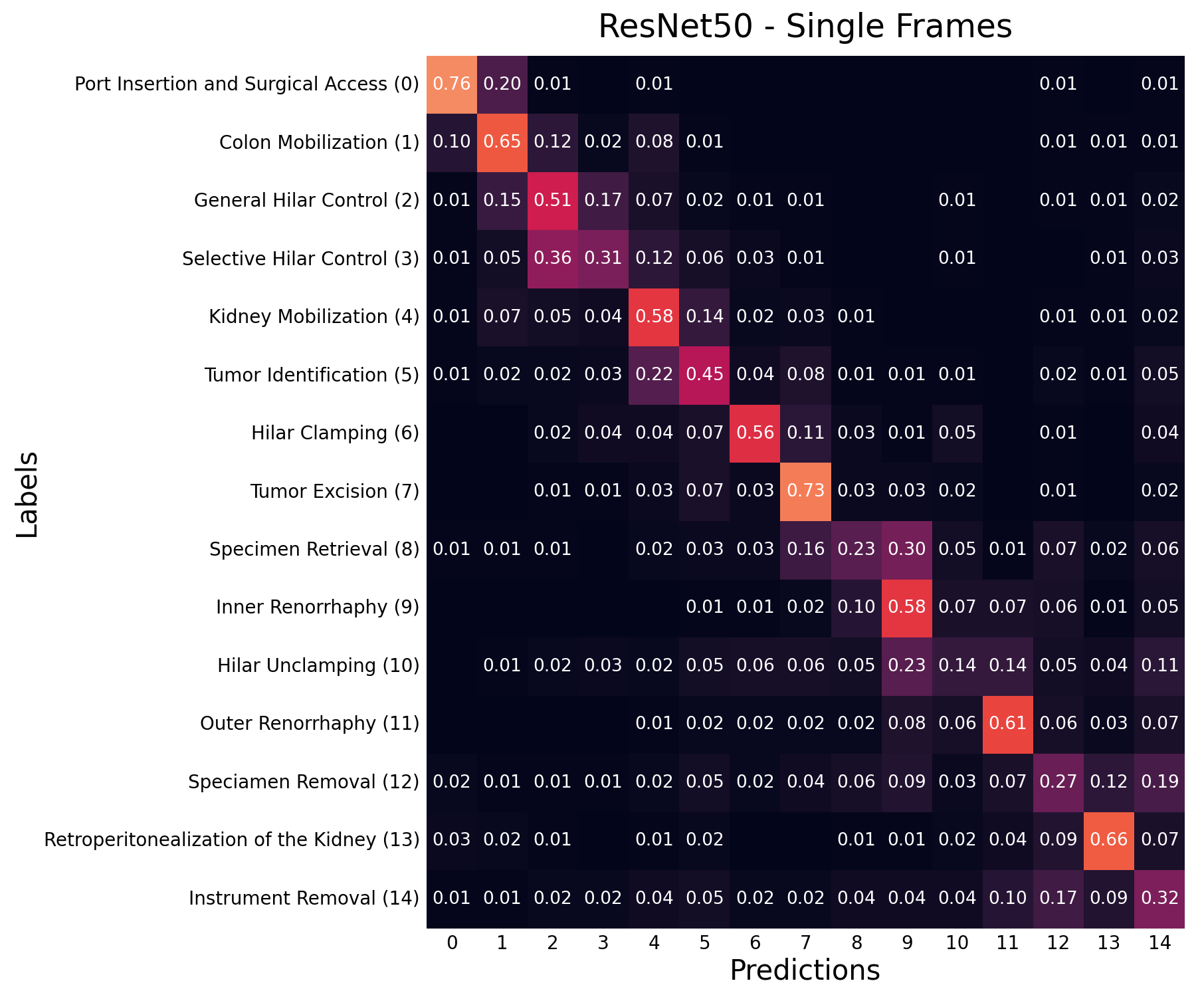}
        \caption{ResNet50 performance for single frames classification}
        \label{fig:confusion_matrix_AI_frames}
     \end{subfigure}
     \begin{subfigure}[b]{1\textwidth}
        \centering
        \includegraphics[width=1\linewidth]{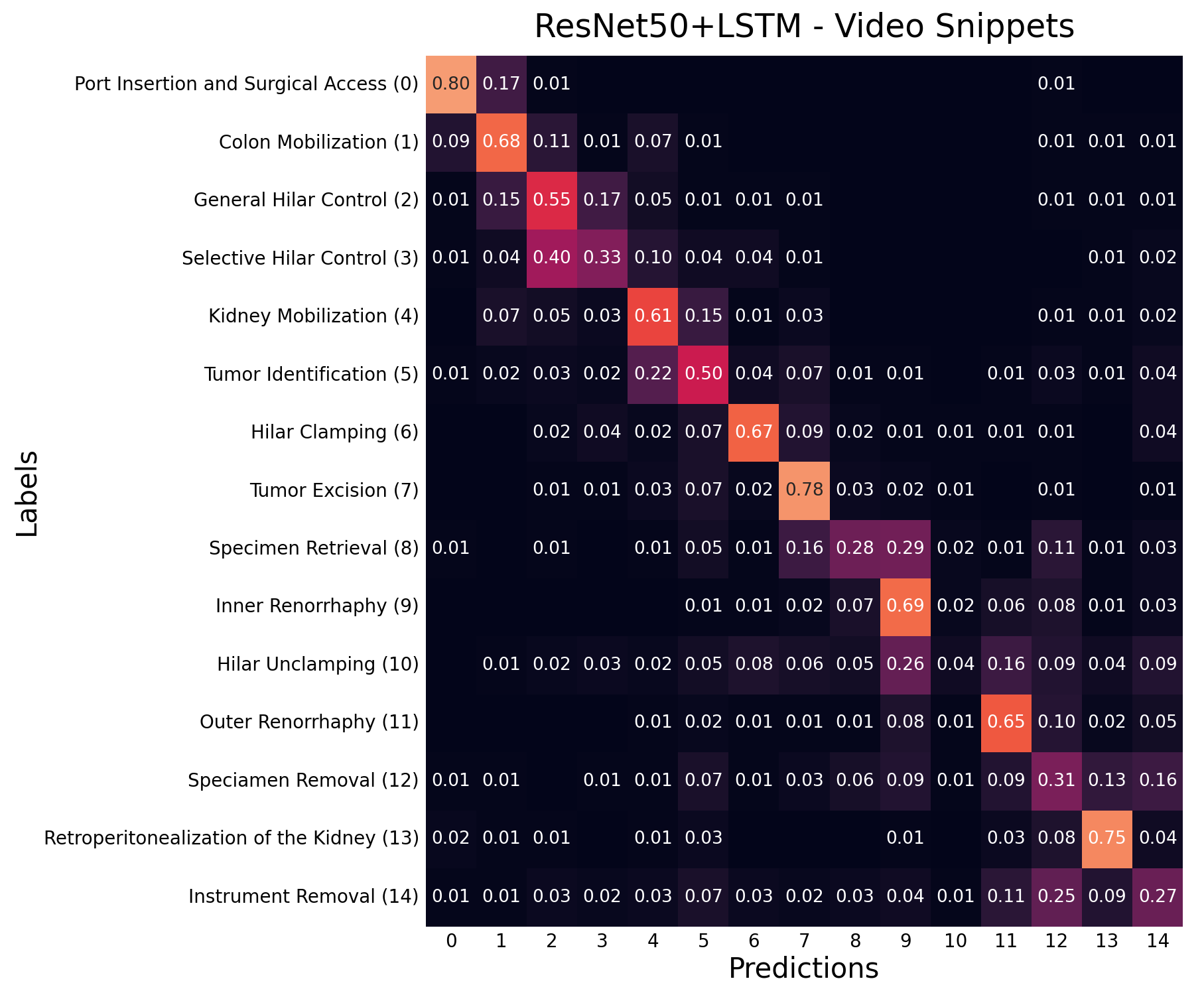}
        \caption{ResNet50+LSTM performance for 10-seconds video snippets classification}
        \label{fig:confusion_matrix_AI_snippets}
     \end{subfigure}
\end{figure}

\begin{figure}[H]
    \ContinuedFloat
    \begin{subfigure}[b]{1\textwidth}
        \centering
        \includegraphics[width=1\linewidth]{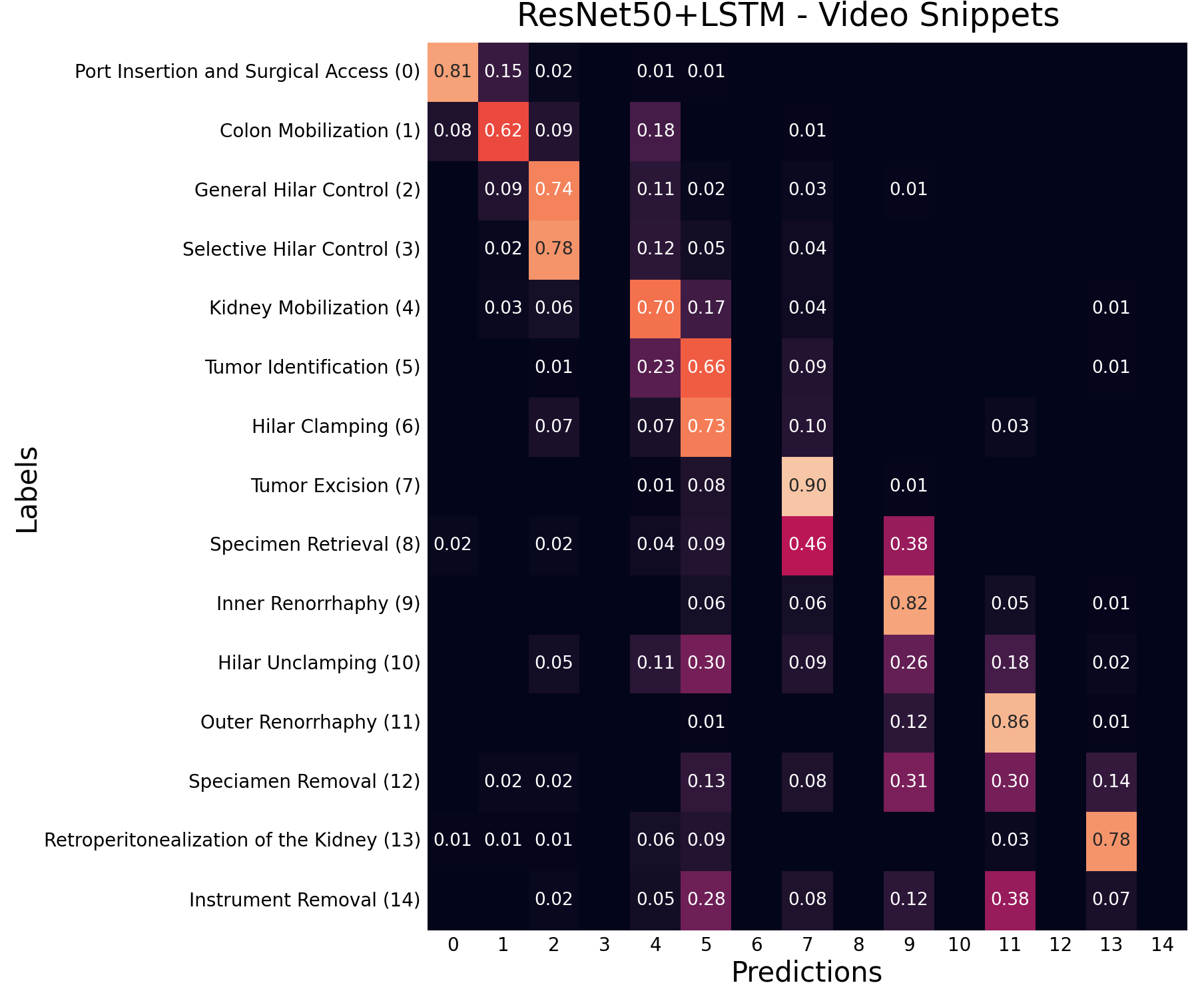}
        \caption{ResNet50+LSTM performance for 60-seconds video snippets classification}
        \label{fig:confusion_matrix_AI_snippets_60_seconds}
     \end{subfigure}
     \begin{subfigure}[b]{1\textwidth}
        \centering
        \includegraphics[width=1\linewidth]{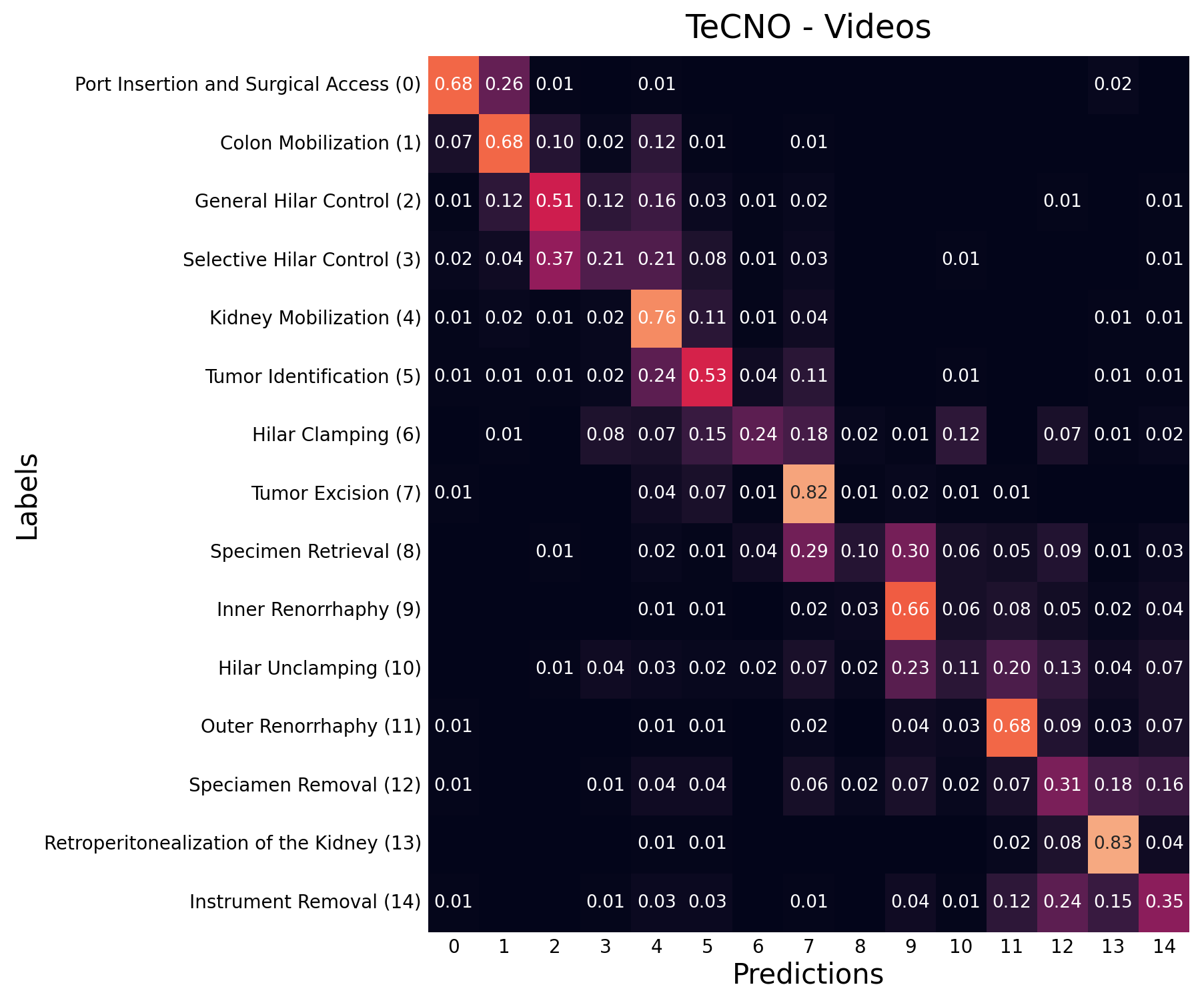}
        \caption{TeCNO performance on phase classification}
        \label{fig:confusion_matrix_AI_snippets_tecno}
     \end{subfigure}
     \caption{Confusion matrices normalised over the predictions of AI models trained over 100\% RAPN dataset.}
     \label{fig:confustion_matrices_AI}
\end{figure}

\begin{figure}[H]
    \begin{subfigure}[b]{1\textwidth}
     \centering
     \begin{subfigure}[b]{0.31\textwidth}
         \centering
         \includegraphics[scale=0.12]{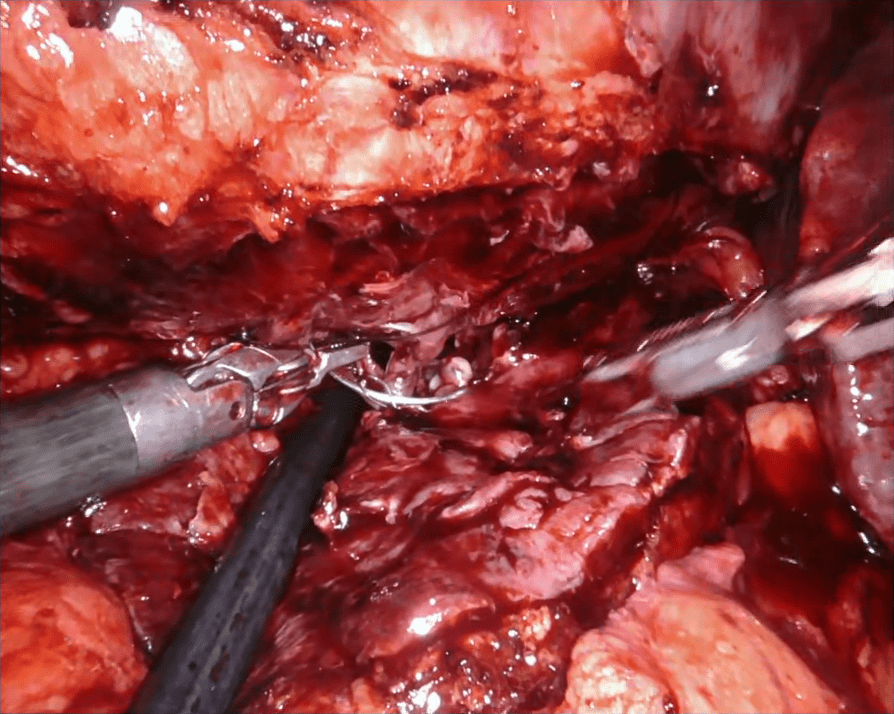}
     \end{subfigure}
     \begin{subfigure}[b]{0.31\textwidth}
         \centering
         \includegraphics[scale=0.085]{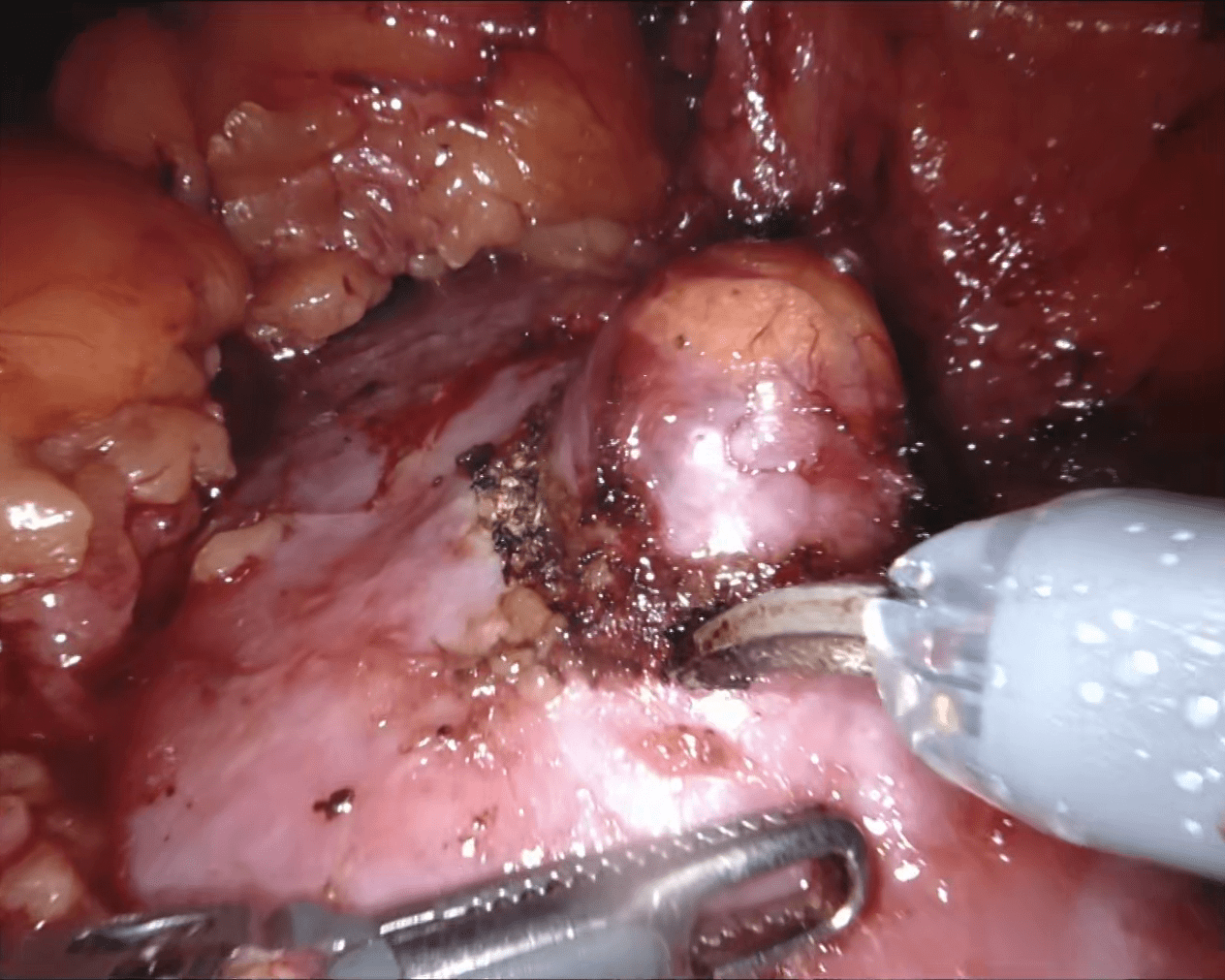}
     \end{subfigure}
     \begin{subfigure}[b]{0.31\textwidth}
         \centering
         \includegraphics[scale=0.085]{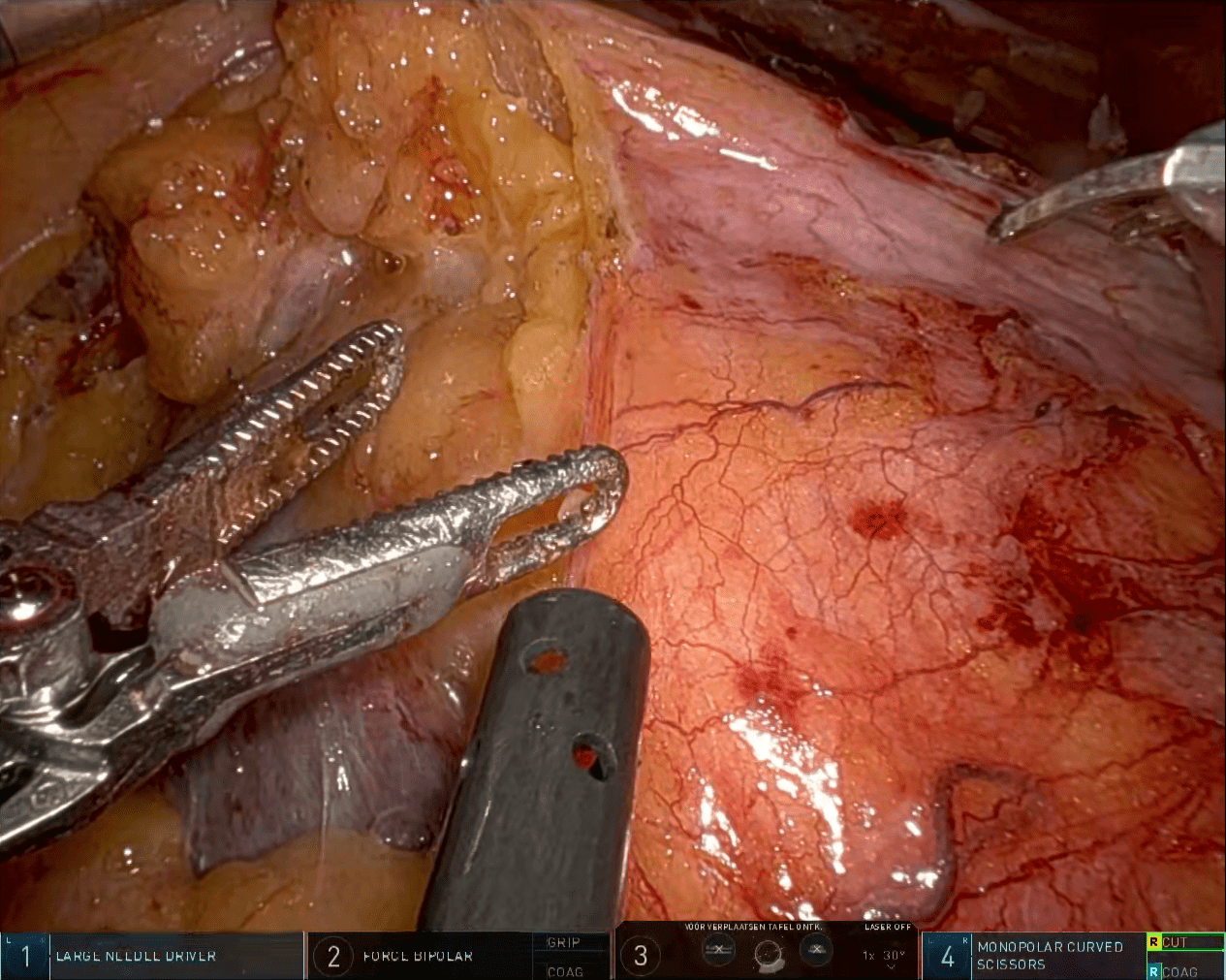}
     \end{subfigure}
    \caption{Clear single frames examples.}
    \label{fig:good_pictures}
    \end{subfigure}
     \begin{subfigure}[b]{1\textwidth}
        \centering
         \begin{subfigure}[b]{0.31\textwidth}
             \centering
             \includegraphics[scale=0.085]{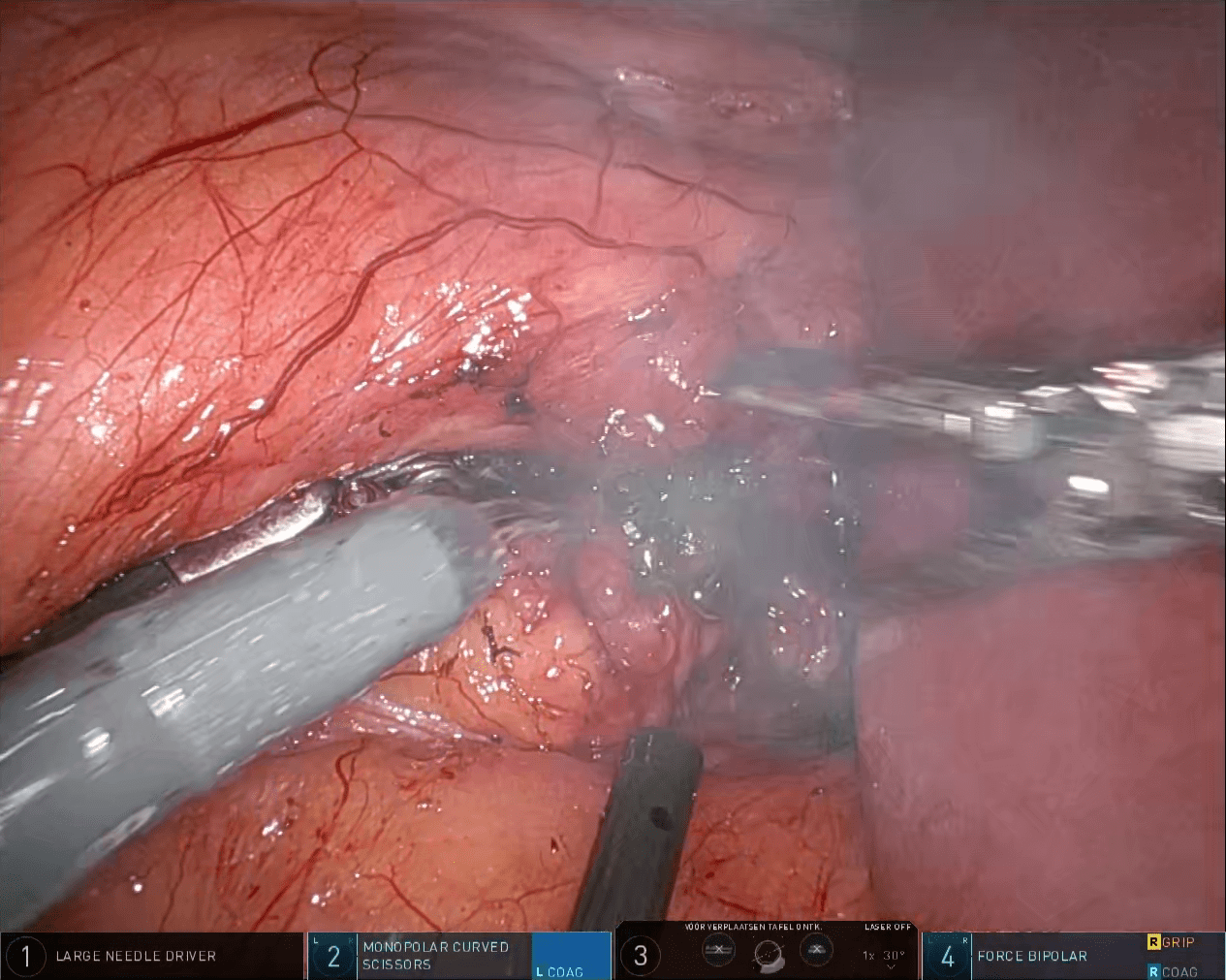}
         \end{subfigure}
         \begin{subfigure}[b]{0.31\textwidth}
             \centering
             \includegraphics[scale=0.085]{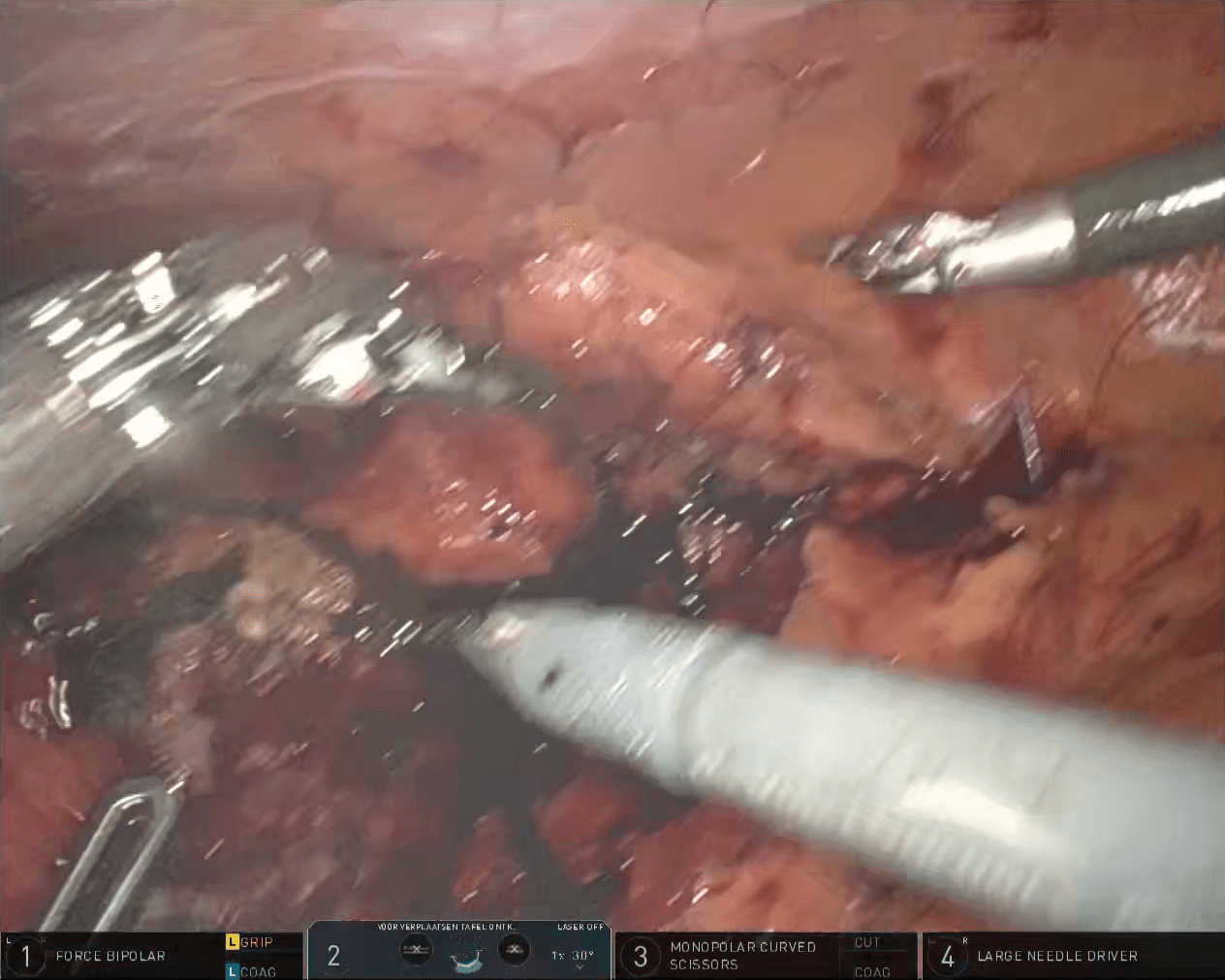}
         \end{subfigure}
         \begin{subfigure}[b]{0.31\textwidth}
             \centering
             \includegraphics[scale=0.085]{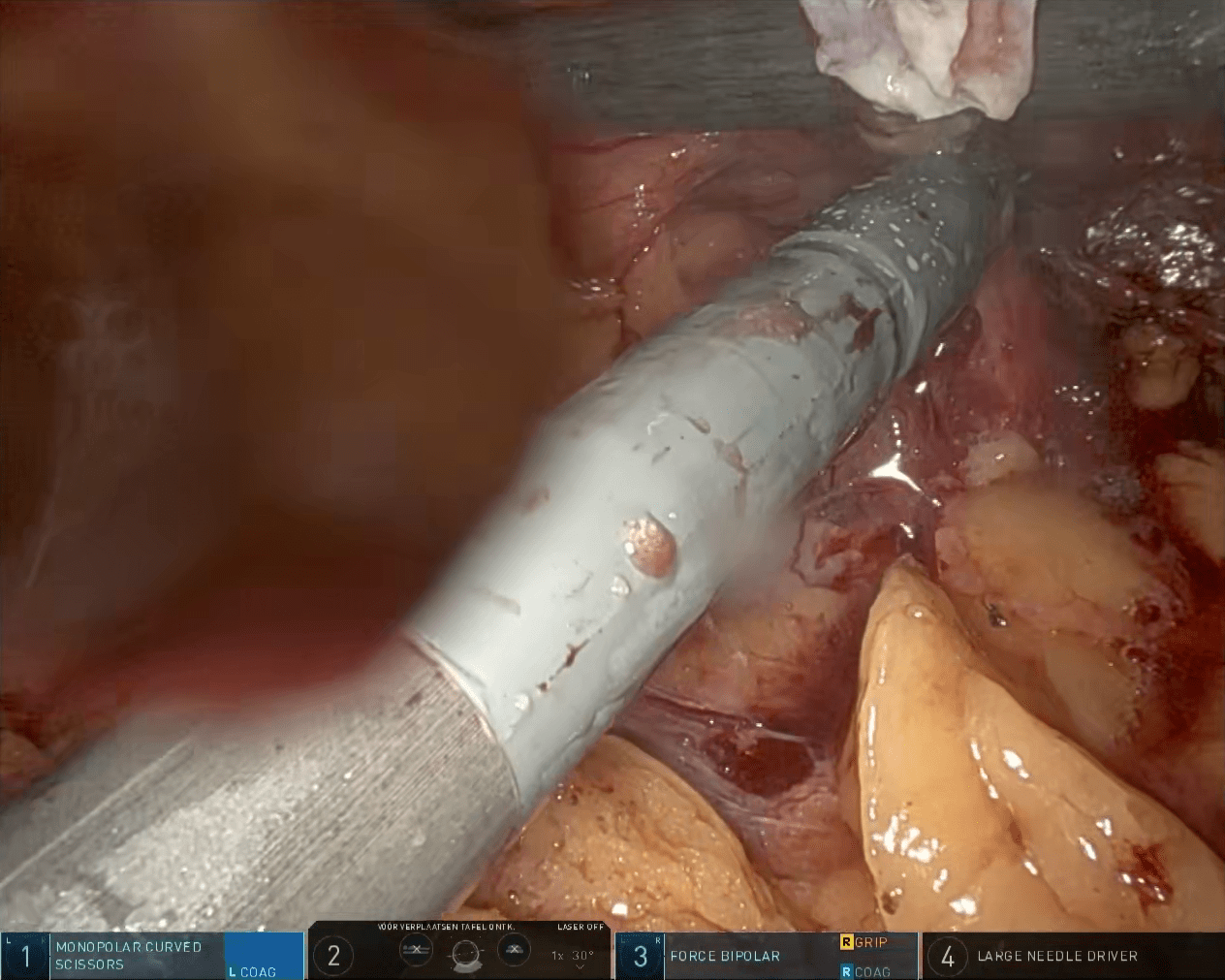}
        \end{subfigure}
        \caption{Difficult single frames examples.}
        \label{fig:bad_pictures}
    \end{subfigure}
    \caption{Single Frames sampled from the set of questions proposed to human raters.}
    \label{fig:good_bad_pictures}
\end{figure}




\end{appendices}

\end{document}